\documentclass[A4paper, 11pt, 3p, twocolumn]{elsarticle}

\usepackage{lineno,hyperref}
\usepackage{amsmath,amssymb}
\modulolinenumbers[5]
\journal{Journal of \LaTeX\ Templates}

\linenumbers

\begin{document}

\begin{frontmatter}

\title{BlueSTEAl: A pair of silicon arrays and a zero-degree phoswich detector for studies of scattering and reactions in inverse kinematics}

%% Group authors per affiliation:
\author[address1,address2]{S. Ota}\corref{mycorrespondingauthor}
%\ead[url]{sota@bnl.gov}
%\author[mysecondaryaddress]{Global Customer Service}
\cortext[mycorrespondingauthor]{Corresponding author}
\ead{sota@bnl.gov}

\author[address3,address4]{G. Christian}
\author[address5]{B. J. Reed}
\author[address5]{W. N. Catford}
\author[address1,address4]{S. Dede}
\author[address5]{D. T. Doherty}
\author[address5]{G. Lotay}
\author[address1,address4]{M. Roosa}
\author[address1]{A. Saastamoinen}
\author[address1,address4]{D.P. Scriven}

\address[address1]{Cyclotron Institute, Texas A\&M University, College Station, TX 77843, USA}
\address[address2]{National Nuclear Data Center, Brookhaven National Laboratory, Upton, NY 11973-5000, USA}
\address[address3]{Department of Astronomy \& Physics, Saint Mary's University, Halifax, NS B3H~3C3, Canada}
\address[address4]{Department of Physics \& Astronomy, Texas A\&M University, College Station, TX 77843, USA}
\address[address5]{Department of Physics, University of Surrey, Guildford GU2 7XH, UK}

\begin{abstract}
BlueSTEAl, the Blue (aluminum chamber of) Silicon TElescope Arrays for light nuclei, has been developed to study direct reactions in inverse kinematics, as well as scattering and breakup reactions using radioactive ion beams. 
It is a detector system consisting of a pair of annular silicon detector arrays and a zero-degree phoswich plastic scintillator. 
For typical binary reaction studies in inverse kinematics, light ions are detected by the Si array in coincidence with heavy recoils detected by the phoswich placed at the focal-plane of a zero-degree magnetic spectrometer. The Si array can also be used to detect light nuclei such as berylium and carbon with clear isotope separation, while the phoswich can also be placed at zero degrees without a spectrometer and used as a high-efficiency beam counting monitor with particle identification capability at the rate of up to $\sim$5$\times$10$^{4}$ particles per second. This paper reports on the capabilities of BlueSTEAl as determined by recent experiments performed at the Texas A\&M Cyclotron Institute. The device is also anticipated to be used in future experiments at other radioactive ion beam facilities.
\end{abstract}

\begin{keyword}
\texttt{elsarticle.cls}\sep \LaTeX\sep Elsevier \sep template
\MSC[2010] 00-01\sep  99-00
\end{keyword}

\end{frontmatter}

\linenumbers

\section{Introduction}

%\paragraph{Installation} If the document class \emph{elsarticle} is not available on your computer, you can download and install the system package \emph{texlive-publishers} (Linux) or install the \LaTeX\ package \emph{elsarticle} using the package manager of your \TeX\ installation, which is typically \TeX\ Live or Mik\TeX.

Direct reactions such as nucleon(s)-transfer/pick-up reactions have been used as a powerful tool to study low-lying excited states of various isotopes for many decades. 
Because of the simple nuclear structure of light ions ($p, d, t, $$^{3}$He, and $\alpha$), 
transfer/pick-up reactions such as ($d,p$), ($p,d$), and $(p,t)$ were popularly used to investigate the low-lying excited states by determining their energy, spin-parity, and spectroscopic factors via a binary direct reaction (e.g., \cite{Ota2015,Ota2021}). 
The $(^{6}$Li,$d$) and $(^{7}$Li,$t$) reactions are also commonly used because of their selectivity for $\alpha$-cluster states \cite{Ota2021,Fulbright1979}. 
For many years, the majority of such transfer/pick-up experiments were performed using light-ion beams with a magnetic spectrometer as a momentum analyzer of light ejecta in normal kinematics. 
The excellent energy--resolution (5--15 keV in FWHM) achieved by the magnetic spectrometers enabled the identification of the populated states uniquely, combined with measurements of the angular distributions \cite{Ball1968,Pringle1986,Lindner1988,Marshall2019}. 
Since the advent of radioactive ion beam technologies, transfer/pick-up reactions have become more widely used in inverse kinematics to study the structure of neutron/proton-rich unstable nuclei \cite{Davids2003a,Catford2010,Jones2010,Pain2015}. 
%It was a turning point of transfer reaction studies because such reactions occur in inverse kinematics and light ejecta are broadly distributed. 
In such inverse kinematics experiments, particularly for neutron-rich isotopes, measurements of a light ejectile's spectrum to deduce the excitation energies ($E_x$) of the populated states are primarily performed at backward angles (corresponding to forward angles in normal kinematics) in the laboratory system. 
Thus, the measurements of light ejecta's energies and angular distributions typically rely on high efficiency (large solid angle coverage) and highly-segmented Si detectors placed at backward angles instead of the magnetic spectrometers. 
Such Si detector systems naturally involve hundreds to thousands readout channels to cover the large angular range with high angular resolution to minimize the uncertainties in deduced $E_x$. 
Indeed, it is a known issue that light ejectile's energy for a given excitation state varies more steeply as a function of angle in inverse kinematics than in normal kinematics \cite{Bardayan2013}, 
which makes angular resolution of the detector a critical matter to achieve good energy resolution in the deduced $E_x$. 
Some of the detector systems add supporting detectors, e.g., high energy-resolution $\gamma$-ray detectors (HPGe and scintillators) around the target position, such as TIARA \cite{Labiche2010}, SHARC \cite{Diget2011}, GODDESS \cite{Pain2017}, and HAGRiD \cite{Smith2018} to compensate for the limited energy resolution of charged-particle measurements with Si detectors. 
This $\gamma$-tagging technique is effective when relatively intense beams or count rates are available. 
Some other detectors such as HELIOS \cite{Wuosmaa2007} and SOLARIS \cite{Solaris2018} involve magnetic fields to achieve good energy resolution without $\gamma$-ray tagging. 
Broadly speaking, in the radioactive ion beam era, experimental studies of binary direct reactions typically involve the development of ever-more complex detectors and associated electronics and data acquisition systems.

Direct reaction studies in inverse kinematics can become more sensitive by detecting the heavy recoil particles at forward angles, typically with a magnetic spectrometer (used as a recoil mass separator), in coincidence with the light ejecta. 
Such coincidence measurements drastically suppress background events in the light ejectile (hence, heavy recoil excitation) spectrum. 
In particular, CD$_2$ and CH$_2$ targets, the most frequently used targets to study $(d,p)$ and $(p,d)$ reactions in inverse kinematics, generate non-negligible contributions of light particles produced by compound nuclear reactions on the carbon component of the target \cite{Margerin2015}. 
Such compound nuclear reaction products can be effectively removed by gating on the heavy recoils resulting from the reaction of interest in a zero-degree detector. For lighter recoils (Z $\lesssim 6$), and measurements with low-intensity radioactive beams ($\lesssim10^4$ particles/second (pps)), zero-degree energy loss vs. total energy signals alone may be sufficient to identify the recoil. However, for heavier recoils and/or more intense beams, an (electro-) magnetic separator is typically required, both to remove the unreacted beam and to improve particle separation and identification. A further advantage of this technique is that for particle-unbound states, the branching ratios of various decay channels can be determined by gating on the resulting heavy recoils. 

Our recent study of the $^{22}$Ne($^6$Li,$d)$$^{26}$Mg reaction using the TIARA Si array \cite{Labiche2010}, combined with the MDM spectrometer at Texas A\&M University \cite{Pringle1986}, is one example of a direct reaction experiment in inverse kinematics, in which  
light ejecta ($d$), heavy recoils ($^{26,25}$Mg), and $\gamma$-rays from the heavy recoils were all measured in coincidence \cite{Ota2020}. 
In the experiment, we successfully demonstrated no background $E_x$ spectra owing to clear selection of $^{25,26}$Mg isotopes with the MDM spectrometer, 
leading to determination of $n$ and $\gamma$ branching ratios. 
Such low background $E_x$ spectra obtained by coincidence techniques, particularly between recoils and ejecta, shed light on some superiorities of inverse kinematics techniques to normal kinematics \cite{Giesen1993,Talwar2016} even for the stable beam experiments. 
We also observed minimal contamination from scattered ${}^{22}$Ne beam in the focal plane detectors owing to non-trivial differences in magnetic rigidity ($B\rho$) between beam and recoil particles, which allowed for the use of the standard TAMU-MDM focal plane detector \cite{Spiridon2019}. The detector uses resistive avalanche wire counters and ionization chambers to measure the positions and energy loss of the incident ions, which limits the focal plane count rate to $\sim$5$\times$10$^3$ pps.

As such, the above-mentioned ($^6$Li,$d$) $\alpha$-transfer experiment was successful. Meanwhile, attempts to apply the technique to other reactions have encountered difficulties with the rates of scattered beam particles into the focal plane detectors. In particular, an early attempt to measure the $^{21}$Ne($p,t$)$^{19}$Ne reaction in inverse kinematics, at a beam energy of 40 MeV/nucleon, suffered from scattered beam in the focal plane at a rate of ~50 counts/pA of primary beam current \cite{Christian2020}. Since the objective of the experiment is to determine the very weak $\alpha$ decay branch of the astrophysically-important $E_x = 4.03$ MeV state in ${}^{19}$Ne \cite{Davids2011}, the ($p,t$) measurement requires beam currents on the order of $\gtrsim$1 nA, resulting in focal-plane count rates of up to 5$\times$10$^4$ pps. Such large rates are not compatible with the traditional TAMU-MDM detector as mentioned above. 
%in particular the position-sensitive resistive-avalanche detectors \cite{Spiridon2019}. 
The TIARA Si detectors were also unsuitable for the ($p,t$) experiment as they are unable to stop the $\sim$40 MeV tritons resulting from the $^{21}$Ne($p,t$) reaction.

Motivated by the issues above, we developed a new detector system to improve and compensate for the various shortages of the TAMU-MDM detector system. 
The improvements are characterized by 
1) a phoswich + double Parallel Plate Avalanche Counter (PPAC) focal plane detector, which provides $\Delta E-E$ and position information while being able to count at incident rates up to 10$^4$--10$^5$ pps; 
%2) It can also measure elastic scattering of light-mass particles with $^{12}$C and ${208}$Pb targets. 
2) a stack of up to four annular Si detectors, which provides $\Delta E-E$ information for the light ejecta and is able to stop high-energy light ejecta scattered at forward angles ($\gtrsim 5^{\circ}$), 
3) replacing the analog data acquisition (DAQ) system with a digital DAQ system developed by Mesytec. The use of a digital DAQ allows for significantly higher count rates than analog systems, due to the resulting short dead time.

In the new focal-plane detector system, 
%our focal-plane detectors consisted of an ionization chamber, proportional wire counters and a thick plastic scintillator. 
%The ionization chamber output was supplemented by a segmented $\mu$-megas \cite{Spiridon2019} to improve the energy resolution, and thus $\Delta E-E$ signals were available combined with the scintillator, or ion chamber pluses $\mu$-megas for low energy heavy ions. 
a phoswich made from EJ-262 fast-response and EJ-240 slow-response plastic scintillator provides the $\Delta E-E$ information replacing the former system with ionization chamber plus a single layer of a thick (0.25$^{\prime\prime}$) plastic scintillator \cite{Spiridon2019}. 
As demonstrated below, the phoswich provides comparable $Z$-identification capability to the latter detectors (see e.g., \cite{Spiridon2019,Ota2020}), but with the ability to withstand approximately an order of magnitude higher incident particle rate. 
High-position resolution ($x$ and $y$ directions) and fast-timing double PPAC detectors \cite{Harris2023} replace the formerly existing proportional wire counters, again with the ability to withstand higher count rates. 
%The new focal-plane detector focus on more high count rates with faster detector responses, while the $\Delta E-E$ signals are still provided by the phoswich. 
In a newly-designed target chamber, we set up a Si $\Delta E-E$ detector constituted of four layers of the S2 Annular Si detectors (one 0.5 mm and three 1.5 mm in thicknesses; Micron Semiconductor Ltd.) \cite{Micron2020}, which can make a 5 mm thick Si $\Delta E-E$ detector system. 
Such thick Si detectors with high segmentation can detect light ejecta at forward angles, which typically have a high energy such as 30--40 MeV. 
It should be noted that there exist some detector systems tailored to measure such high energy light ions and light nuclei. 
However, these detectors, e.g., LASSA \cite{Davin2001}, MUST \cite{Blumenfeld1999}, MUST2 \cite{Pollacco2003}, and HiRA \cite{Wallace2007}, employ CsI scintillators as the residual-energy ($E_{res}$) detector, 
and therefore the energy resolution in the reconstructed heavy recoil's $E_x$ spectrum suffers due to the relatively poor energy resolution of CsI as compared to silicon. In particular, for the $^{21}$Ne($p,t$) experiment, GEANT4 \cite{Agostinell2003} simulations showed that using a CsI detector would make it impossible to separate the 4.03 MeV state of interest from neighboring states at higher energy \cite{Dede2019}. These higher-energy states have larger $\alpha$ branching ratios and therefore it would cost us if we use CsI. 
A new digital DAQ system is also introduced, in which output pulses from photomultipliers (PMTs) and preamplifiers (preamps) are fully digitized and processed with FPGA (Field-Programmable Gate Array) implemented in Mesytec VME (Versatile Module Europe) modules. 
The new DAQ system significantly reduces the dead time of our measurements from the order of a few tens to hundreds $\mu$s to about 1 $\mu$s per event.
Together with the phoswich's excellent timing resolution, it is now possible to perform experiments in high beam flux environments. 
%and also for precise absolute elastic and breakup cross section measurements. 

As described below, our new thick Si detector system was tested with a study of the $^{21}$Ne$(p,t)$ reaction. 
%, in which typically $t$ are emitted at forward angles with high energy (e.g., 40 MeV) because of the high-energy beam to overcome the negative reaction Q-values \cite{Davids2003}. 
The detector setup, however, can also be split into two sets of a $\Delta E-E$ detector (e.g., 0.5 mm+1.5 mm and 2$\times$1.5 mm), being placed at small and large angles, respectively. 
The latter setup enables us to measure angular distributions efficiently in one single experiment, which is important when the beam intensity and statistics are limited. 
We employed the set up for the measurements of elastic scattering and neutron breakup reactions of $^{11}$Be in one of our recent experiments. 
%The scattering and breakup measurements need good energy resolution as well to clearly separate breakup events from scattering events by isotope identification.
In both setups, the Si detectors are housed in a newly designed vacuum chamber, BlueSTEAl, the Blue (aluminum chamber of) Silicon TElescope Arrays for light nuclei. 
The new Si Detector arrays are also compatible with the TIARA scattering chamber, e.g. being used to cover forward scattering angles in experiments employing coincident $\gamma$-ray detection (the TIARA chamber has a thin neck at the target position to facilitate placement of HPGe detectors surrounding the target at the minimum distance). However, we developed the BlueSTEAl chamber for more convenient and flexible use for experiments not employing $\gamma$-ray detectors.

%In the third point, we recently developed a new research program to measure elastic and neutron beakup cross sections of halo nuclei.
%The experiments need clear separation of the halo nuclei (A) and breakup product (A-1). 
%Since it involves heavy targets, the kinematics follow normal kinematics (sharp forward angle scattering). 
%In this case, detection of beam particles are still important to monitor and normalize the total beam counts to determine absolute elastic and breakup cross sections. 

In the following, we describe the design and capability of our new detector system, BlueSTEAl. 
The BlueSTEAl also contains some supporting detectors (Si surface barrier detectors and diamond detectors) and 
high-utility flanges based on a chamber developed by Ideal Vacuum Products, LLC \cite{IdealVacuum2019}. 
The Si detector efficiency in the BlueSteAl chamber is evaluated with Monte Carlo simulations in detail. 
Finally, we briefly introduce the preliminary results from new measurements of $^{21}$Ne$(p,t)$ reaction and $^{12}$C($^{11}$Be,$^{11}$Be/$^{10}$Be) elastic scattering / neutron breakup with BlueSTEAl. 
Future possibilities of performing experiments with BlueSTEAl using rare isotope beams, e.g., at FRIB (Facility for Rare Isotope Beams) are also discussed.

%\paragraph{Usage} Once the package is properly installed, you can use the document class \emph{elsarticle} to create a manuscript. Please make sure that your manuscript follows the guidelines in the Guide for Authors of the relevant journal. It is not necessary to typeset your manuscript in exactly the same way as an article, unless you are submitting to a camera-ready copy (CRC) journal.

\section{Design description of detector components}
\subsection{BlueSTEAl scattering chamber}

The BlueSteAl scattering chamber was designed taking advantage of a commercially available cubic modular vacuum chamber from Ideal Vacuum Product, LLC. 
The chamber is composed of a cubic frame with the dimensions of 12$^{\prime\prime}$$\times$12$^{\prime\prime}$$\times$12$^{\prime\prime}$ (L$\times$W$\times$H) and flanges to seal the chamber. 
All components are made from 6061-T6 Aluminum Alloy. 
We selected the chamber because of the convenient geometrical size to house Si detectors and accessibility to the inside of the chamber. 
For instance, two sizes of flanges with 12$^{\prime\prime}$$\times$6$^{\prime\prime}$ (L$\times$W) and 12$^{\prime\prime}$$\times$12$^{\prime\prime}$ (both thicknesses are 0.25$^{\prime\prime}$) are interchangeably used to seal the chamber frame. 
Both flanges can accommodate various external ports such as ISO100/160 (12$^{\prime\prime}$$\times$12$^{\prime\prime}$ flange only), KF-40, and CF 4.5. 
We set up BlueSTEAl by closing up with these two types of flanges depending on the experimental requirement and convenience. 
%It gives us flexible choices on how we connect the chamber to beam pipes, vacuum pumps, external cameras, and feedthrough. 
In the most common setup, we used the six 12$^{\prime\prime}$$\times$6$^{\prime\prime}$ flanges for the three sides (left, right and top in beam-view (seen from the upstream side); two flanges per side) and used the three 12$^{\prime\prime}$$\times$12$^{\prime\prime}$ flanges for the remaining three sides of the chamber cube. 
Each 12$^{\prime\prime}$$\times$6$^{\prime\prime}$ flange accommodates three KF-40 ports for accessory parts (a vacuum gauge, target ladders, and so on) or a 6.9$^{\prime\prime}$$\times$4.3$^{\prime\prime}$ (0.2$^{\prime\prime}$ thickness) printed circuit board (PCB) to connect Si detector cables to external preamps. 
The two 12$^{\prime\prime}$$\times$12$^{\prime\prime}$ flanges placed on the beam upstream and downstream sides have an ISO-100/160 port to couple to beam pipes, and the one last blank 12$^{\prime\prime}$$\times$12$^{\prime\prime}$ flange is attached to  the bottom side of the chamber cube to place the Si detectors on it. 
While the beam pipe attached to either (upstream or downstream) end is stable enough to support the chamber ($<$100 lbs, including all detectors, accessory parts, and electronics), 
we sit BlueSTEAl on a laboratory jack (McMaster-Carr, 9967T45) to position the chamber to be coupled with the beam pipes providing extra stability. 

\begin{figure*}[!ht]
        \centering
          \includegraphics[width=16cm]{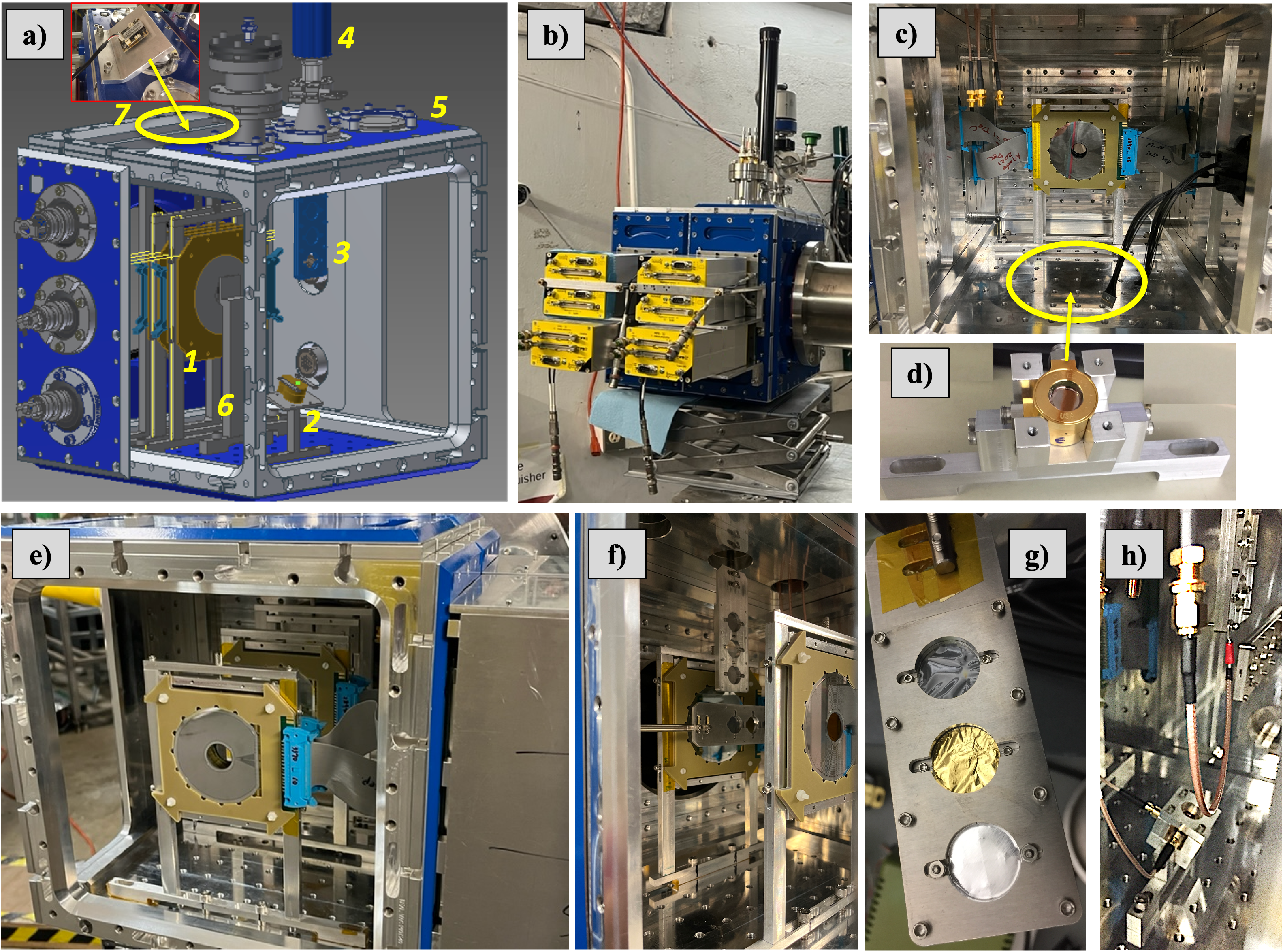}          
        \caption{CAD (Computer Aided Design) design and photographs of the BlueSTEAl detectors. a) Illustration of the BlueSTEAl chamber. A beam is directed from right to left, $\it{1}$ Si annular (Micron S2) detectors attached to the position-adjustable detector holders, $\it{2}$ Si surface barrier detectors, $\it{3}$ target ladder, $\it{4}$ target actuator, $\it{5}$ 6$^{\prime\prime}$$\times$12$^{\prime\prime}$ flange (with one blank and one with a feedthrough KF40 port), $\it{6}$ $\alpha$ source stand, $\it{7}$ the external web camera to monitor the beam spot, b) the BlueSTEAl chamber with preamps mounted (used for the four Si S2), c) inside the chamber (beam upstream view), d) Si surface barrier detectors ($\Delta E-E$) in the detector holder (to be placed in the marked position in c)), e) inside the chamber (downstream view), f) two target ladders set from the top and the side directions, g) target ladder with targets mounted, h) Si surface barrier detectors angled toward the target position. 
%\cite{Glatz1986}. %
        }\label{fig:Fig1}
%    \end{minipage}
    \hfill{}
\end{figure*}

\begin{figure}[!ht]
        \centering
          \includegraphics[width=8cm]{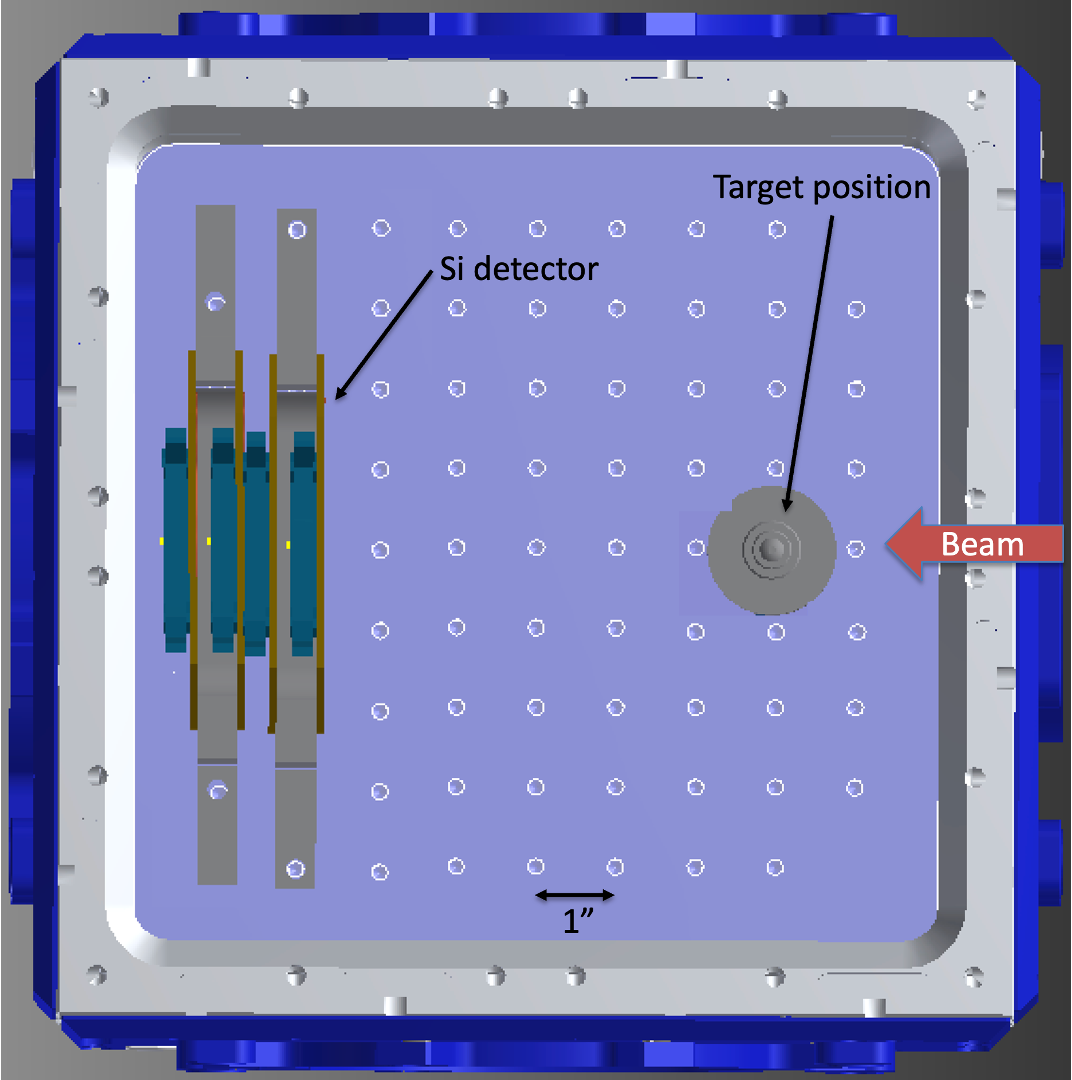}          
        \caption{(left panel) Screw hole mapping in the 12$^{\prime\prime}$$\times$12$^{\prime\prime}$ floor flange of the BlueSTEAl chamber (view from the top of the chamber). Si detectors are attached via the detector holder with two screw holes. Because the holes are threaded every 1$^{\prime\prime}$ interval, the Si detectors are typically distanced downstream from the target by one of 1$^{\prime\prime}$, 2$^{\prime\prime}$, 3$^{\prime\prime}$, 4$^{\prime\prime}$, 5$^{\prime\prime}$, 6$^{\prime\prime}$, and 7$^{\prime\prime}$ distances. 
%\cite{Glatz1986}. %
        }\label{fig:Fig2a}
%    \end{minipage}
    \hfill{}
\end{figure}

\begin{figure}[!ht]
        \centering
          \includegraphics[width=8cm]{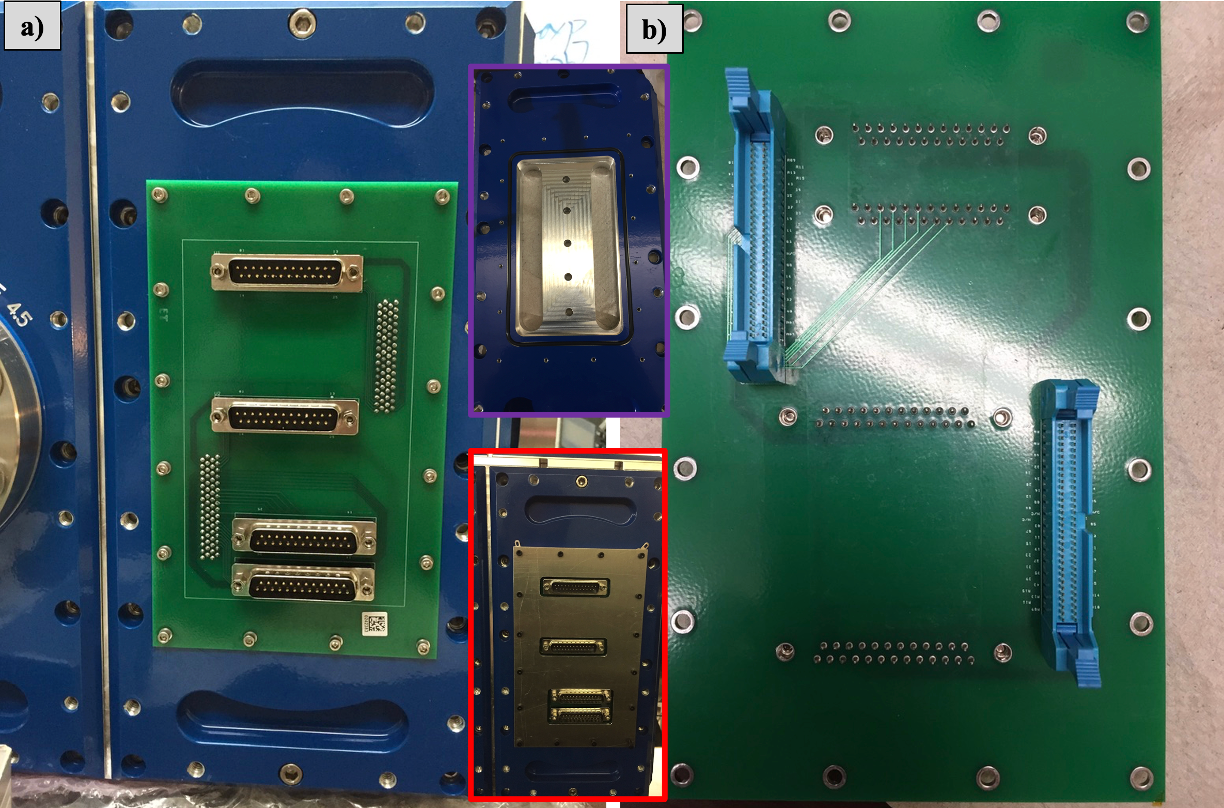}          
        \caption{a) A PCB board for Si preamps (to accommodate two MPR-16s and a MPR-32) attached to a 6$^{\prime\prime}$$\times$12$^{\prime\prime}$ flange. The o-ring groove is cut on air side (the photograph enclosed by purple line). In the actual setting, as shown in the photograph (enclosed by red line), the board is covered with an aluminum frame (0.188$^{\prime\prime}$ thickness) to reduce the electrical noise by completing the Faraday Cage. b) the vacuum side of the same PCB. 
%\cite{Glatz1986}. %
        }\label{fig:Fig2}
%    \end{minipage}
    \hfill{}
\end{figure}

%or used to attach a phoswich scintillator as a beam dump as described below. 
The vacuum side of each flange has an array of 1/4$^{\prime\prime}$-20 screw threaded holes every 1$^{\prime\prime}$ interval. 
These holes are conveniently used to fix detectors and $\alpha$-source holders inside the chamber, particularly on the bottom 12$^{\prime\prime}$$\times$12$^{\prime\prime}$ flange. 
Figure~\ref{fig:Fig1} shows our typical configuration of BlueSTEAl using these flanges. 
The figures demonstrate the accessibility to the inner side of the chamber. 
Because each flange is easily removable with no interference with other flanges, the accessibility from every direction allows one to easily plug and unplug cables as well as mount and unmount detectors. 
Figure~\ref{fig:Fig2a} shows a map of the screw hole array on the 12$^{\prime\prime}$$\times$12$^{\prime\prime}$ flange. 
The Si detectors are attached to two of the holes via the detector holder. 
Thus, we usually set the detectors somewhere between 1$^{\prime\prime}$ and 7$^{\prime\prime}$ (farthest) downstream from the target. 
The PCBs attached to the side flanges shown in Figure~\ref{fig:Fig2} are also designed for sealing a vacuum ($\lesssim10^{-6}$ torr) inside the chamber with a viton o-ring (size 161). 
The o-ring grooves are cut on the 12$^{\prime\prime}$$\times$6$^{\prime\prime}$ flange (air side) (see Figure~\ref{fig:Fig2}). 
Since the PCB designs are easily modifiable, the design of feedthroughs for the Si or other detectors placed inside the chamber is also modifiable. 
In the current setup, we read 23 rings (the neighboring two rings are paired) from the two Si detectors, and 16 sectors from each of the four Si detectors. 
Thus the total number of readout channels from the four Si detectors is currently 110 (23$\times$2+16$\times$4). 
Inside the PCB, Si output signals from the detector's connector (NFP-64A-0314BF, Yamaichi Electronics) are converted to Dsub-25 connector outputs which are an input format of the following Mesytec preamps. 
The sector (ring) signals are fed to Mesytec MPR-16 (32) preamps with adjustable gains (sensitivity) of 5--1500 MeV directly attached to the PCB on the outside of the chamber (see Figure~\ref{fig:Fig1} b)). 

\begin{figure}[!ht]
        \centering
          \includegraphics[width=8cm]{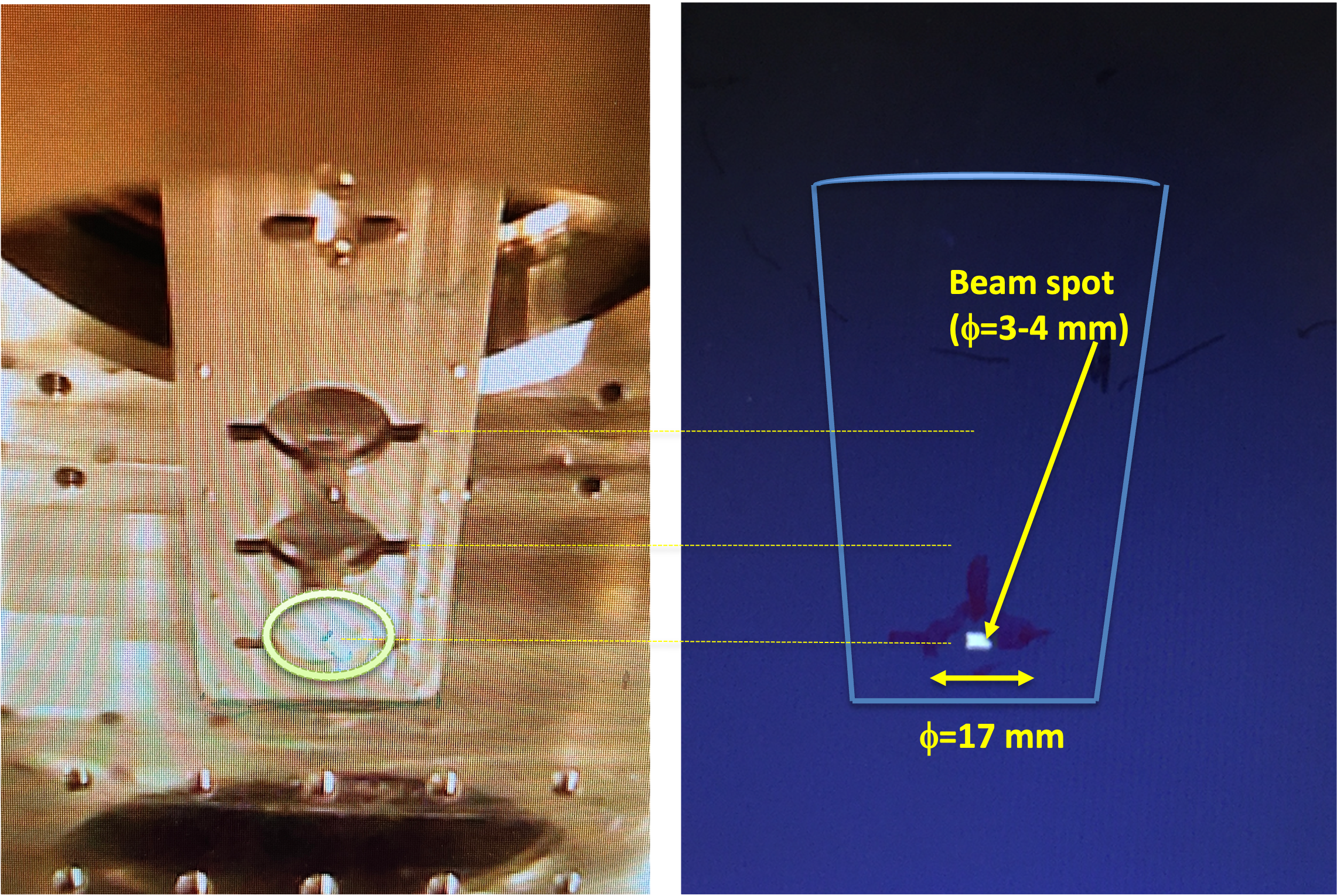}          
        \caption{(left panel) Target ladder monitored from the external web camera. The target set on the beam line is marked by a green circle. (right) The experimental cave's light was turned off so that the camera can observe the illuminated light by the $^{21}$Ne 40 MeV/u beam. The beam spot size was estimated to be $\phi$=3--4 mm, hence the rms $<$ 1 mm (observed with an attenuated (by a factor of 1/100--1/1000) beam). 
%\cite{Glatz1986}. %
        }\label{fig:Fig3}
%    \end{minipage}
    \hfill{}
\end{figure}

The target ladders attached to a manual actuator (6$^{\prime\prime}$ or 4$^{\prime\prime}$ length Manual Linear Motion Feedthrough (L2212-6-SF and L2111-4-SF), Huntington Vacuum Products), vacuum gauges, Lemo, BNC, and SHV feedthroughs, and glass windows (to monitor the inside of the chamber from outside) are all attached to the chamber through one of the KF-40 ports. 
We used two target ladders each of which can accommodate up to three targets (a target's frame size: 1.2$^{\prime\prime}$$\times$0.9$^{\prime\prime}$$\times$(0.033$^{\prime\prime}$$\times$2; sandwiched by two aluminum frames) with a $\phi$=0.67$^{\prime\prime}$ ($\sim$17 mm) hole where a beam hits at the center). 
While the tail of a relatively spread beam such as radioactive ion beams could be blocked and the beam intensity may be reduced due to the hole size, there is also a benefit to limit the size of the beam incident on the target, leading to better angular resolution (see Section 3 and 4). 
Usually, a sheet of phosphor or Zinc Cadmium Sulfide pasted on a thin (0.1 milli inch) aluminum foil to monitor the beam position with the external camera accounts for one of the target positions (Figure~\ref{fig:Fig3}). 
One target ladder is operated from the top of the chamber (along $y$-direction) and the other is from the left or right side of the chamber (along $x$-direction). 
With the linear actuator, the target position in the direction of the operation ($x$ or $y$) is adjustable with sub-millimeter accuracy. 
%Thus targets are manually movable in both X and Y directions, and . 
In a usual case, the target's ($x$,$y$) position is set aligned with detectors along the predefined beam optics line ($z$-direction) within the accuracy of a mm or less with a transit prior to the experiment. 
The phosphor foil on one target ladder is used as a reference point to the other ladder when one needs to replace the targets during the experiment and no proper re-alignment process is possible. 
An external web camera (ELP USB with CMOS Camera 2.1 mm Lens 1080P HD) is attached to the chamber to monitor inside the chamber through the borosilicate glass disc window (with the 2$^{\prime\prime}$ diameter ($\phi$) and 3/16$^{\prime\prime}$ thickness, McMaster-Carr) (Figure~\ref{fig:Fig1} a)). 
A vacuum gauge (PKR251 with a TPG361 controller, Phiffer Vacuum) is attached to one of the KF-40 ports with a small manual venting valve from where the dry nitrogen flows into when breaking the vacuum. 
One KF-40 flange attached to the chamber accommodates four Lemo feedthroughs to read out the signals from beam monitoring detectors (see below) and to bias the target ladder (made of aluminum) to collect the $\delta$-rays expelled from the target by impinging beam (typically biased by 100--300 V). 
KF-40 flanges with BNC and SHV feedthroughs are optionally used to host detectors which need a high voltage such as small scintillators (e.g., p-terphenyl \cite{Scriven2021}). 
One 12$^{\prime\prime}$$\times$6$^{\prime\prime}$ flange which accommodates a CF 4.5$^{\prime\prime}$ feedthrough with two Dsub-25 connectors is also used to read out the diamond $\Delta E-E$ detector signals as described below. 

\subsection{Annular silicon detectors}

The main detectors placed inside the BlueSTEAl chamber are the four Micron S2 annular Si detectors ($\phi$=70 mm active area with $\phi$=20 mm inner hole for the beam to pass through) (Figure~\ref{fig:Fig1} e)). 
The detectors are used to measure light ion energies to deduce the excited levels of heavy recoil nuclei populated by binary reactions in inverse kinematics experiments. 
They are also used to measure energy and angular distributions of elastic$+$inelastic/breakup recoils from experiments using radioactive ion beams. 
The high energy-resolution and angular-resolution are therefore important to fulfill the experiments' requirements. 
Each of our S2 detectors has 16 segments (sectors) in the $\phi$ direction (Ohmic side), 
and 45 segments (rings) in the $\theta$ direction (Junction side). 
The detectors originally have 48 rings each but the first 3 rings are not electrically read out (primarily to avoid detecting the beam scattering at very forward angles). 
The first active ring (4th ring) is read as a single strip, and the following 44 rings are paired by neighboring two rings and electrically conducted in the PCB, forming 23 ring channels in total (in the active area of $\phi$=26.0--70.0 mm). 
Note that the ring configuration is easily switched to the original (45 ring) setting by modifying the PCB. 
Three of these detectors have a 1.5 mm thickness and the other one has a 0.5 mm thickness, respectively. 
Each of the two Si detectors is attached to either side of a detector holder (3/16$^{\prime\prime}$ thickness) made of aluminum, with height adjustability (see Figure~\ref{fig:Fig1} (e)). 
Unshielded high-density cables (60 pins; Flat Cables 30AWG .025$^{\prime\prime}$, 3M) were plugged between the detectors and the PCBs inside the chamber. 
The cable length was kept minimum ($<$10 cm) to reduce electrical noise. 
The output cables are fed to the preamps directly attached to the outside (air side) of the chamber through the PCBs (see Figure~\ref{fig:Fig2}). 
Depending on the experimental needs, the two sets of the Si detector pairs are 
stacked to form a thick (5 mm) $\Delta E-E$ detector, 
%which is effective when we study high energy light particles such as $(p,t)$ reactions. 
or placed separately so that each set of the Si detector pair can cover the near-side (e.g., 2$^{\prime\prime}$ away) and far-side (7$^{\prime\prime}$ away) angles from the target position, respectively. 
In the latter setup, the detectors cover the large angular range of 5--35$^\circ$ with negligible overlap between the two sets of $\Delta E-E$ detectors (see Section 3). 
%This setting was used for elastic scattering measurements. 
%Mesytec preamps with various range of 5--15000 MeV are attached to the PCB directly. 
%

The typical energy resolution of the Si detectors in this setup tested with a commercially available multi-nuclide ($^{244}$Cm, $^{241}$Am, and $^{239}$Pu) $\alpha$ sources ($\sim$5.5 MeV) is 40 keV Full Width at Half Maximum (FWHM). 
When we need to measure higher energy charged-particles, we use an external pulser (Model PB-5 precision NIM pulse generator, Berkeley Nucleonic Corp) to feed into the preamp and check the linearity from the $\alpha$ source energy region to the high energy region at the VME module's ADC outputs \cite{Brown2010}. 
We confirmed the high linearity up to high energy regions such as 140 MeV as studied in $^{11}$Be elastic scattering experiment. 
The energy calibration by this method indeed works well and punch through of some particles are well reproduced as an energy loss calculation code (SRIM2013 \cite{Ziegler2010}) predicts (see Section 4). 

% \begin{figure}[!ht]
%        \centering
%          \includegraphics[width=8cm]{figures/Calibration_Sector_Si3_8.png}          
%        \caption{(top panel) PCB board for Si preamps, (bottom left) the other side of the same PCB (vacuum side), (bottom right) PCB board covered with an aluminum plate for electrical noise reduction. 
%\cite{Glatz1986}. %
%        }\label{fig:Fig5}
%    \end{minipage}
%    \hfill{}
%\end{figure}

\subsection{Zero-degree phoswich scintillator}
\begin{figure*}[!ht]
        \centering
          \includegraphics[width=16cm]{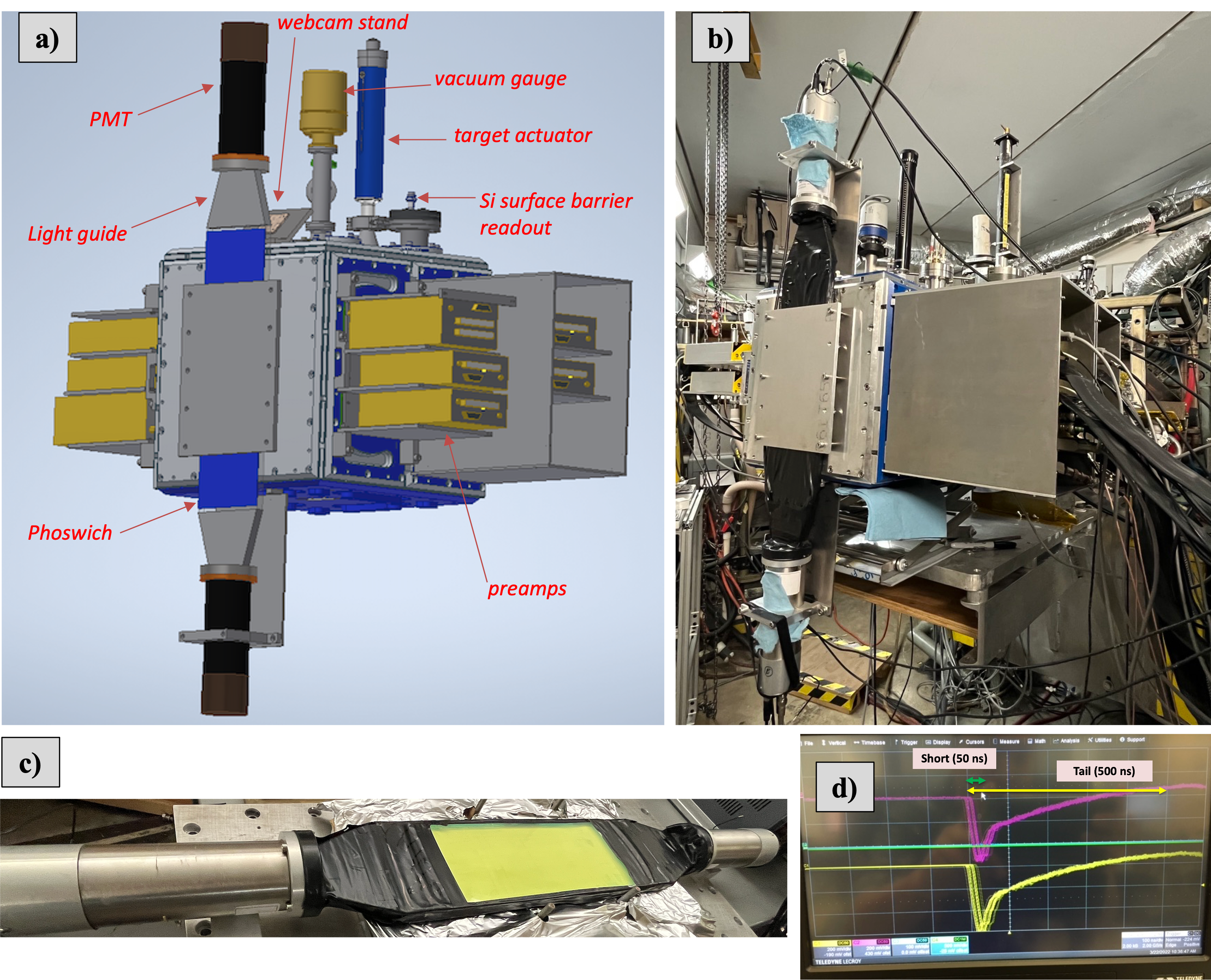}          
        \caption{a) Design drawing of the BlueSTEAl chamber set up for elastic and breakup reaction experiment. b) a photograph taken from the same direction as a), c) phoswich detector front surface (the beam upstream), d) $^{21}$Ne 40 MeV/u beam pulses displayed on the oscilloscope from either end of the PMTs. 
%\cite{Glatz1986}. %
        }\label{fig:Fig4}
%    \end{minipage}
    \hfill{}
\end{figure*}

In inverse kinematics binary reaction experiments, detection of the heavy recoils by clear particle identification in coincidence with the light ions drastically reduces the background in the $E_x$ spectrum. 
For this purpose, a phoswich detector made of two plastic scintillators is placed at the very end of the beam line (at zero degrees). 
The detector at this position must have enough thickness to stop the particles of interest (beam or beam-like recoils) and identify the particle with the $\Delta E-E$ method. Fast response is also required to withstand a high flux of scattered/stray beam with minimum pile-up. 
The detector we selected in the end is a phoswich with a conventional combination of two plastic scintillators, EJ-262 (2.1 ns decay time ($\tau$)) and EJ-240 (285 ns decay time)  \cite{ELJ2020_EJ262,ELJ2020_EJ240}. 
%The size of our phoswich is 10 x 300 cm$^2$ with a light guide coupled to a photomultiplier on either end. 
The fast-decay EJ-262 scintillator primarily emits green light and was specifically chosen to minimize absorption and re-emission within the thick (EJ-240) layer. 
The EJ-262 (front) with a thickness of 1 mm is laminated to the EJ-240 (back) with the thickness of 1 cm (Figure~\ref{fig:Fig4}). 
The dimensions of the EJ-240 and 262 are almost the same except for the thickness and width (37 cm for EJ-262 while 38 cm for EJ-240). 
Both have the 11$\times$7.5 cm trapezoidal shape light guide, 
and the ETEL 9266KB photo-multipliers (PMTs) are shared between the two scintillators and attached to both (left and right) ends. 
Table~\ref{tab:table1} shows the properties of these scintillators. 
The phoswich detector is conveniently used to enable an elemental ($Z$) identification by the $\Delta E-E$ technique from the fast and slow pulses (e.g. \cite{Leegte1992,Fox1996,Davids2003}), and allows one to separate heavy recoil's $Z$ ($>$10) at high count rates. 
Its physical stability in the high count-rate environment of heavy ions also makes the phoswich attractive to use as a focal plane detector for direct reactions. 
In the $^{21}$Ne$(p,t)$ experiment described below, we observed the phoswitch operated at $\sim$5$\times$10$^4$ events / second without significant problems. 
By placing the phoswich downstream from a magnetic spectrometer, the isotopes can be identified from the focal plane positions. 
Although the phoswich can provide the position ($x$,$y$) information from the relative timing differences between the two PMT ends (resolution ($\Delta t$) of $\sim$1.5 ns FWHM typically obtained with our DAQ), more precise position resolution is achieved with additional angular information by using a double PPAC (Parallel Plate Avalanche Counter) detector (timing and position resolution: $\sim$1 ns and $\sim$2 mm FWHM, see \cite{Harris2023} for details). 
Therefore, we used a beam line configuration of MDM spectrometer$\rightarrow$PPAC1$\rightarrow$PPAC2$\rightarrow$phoswich for the $^{21}$Ne$(p,t)$ experiment described below, where each of PPAC1 and 2 consists of a double PPAC to reconstruct $x$ and $y$ positions.

\begin{table*}
%        \centering
  \begin{center}
    \caption{Phoswich scintillator properties \cite{ELJ2020_EJ262,ELJ2020_EJ240}.}
    \label{tab:table1}
%    \begin{ruledtabular}
    \begin{tabular}{ccccc} % <-- Alignments: 1st column left, 2nd middle and 3rd right, with vertical lines in between
          \hline
      $$ & Decay time ($\tau$) &  Efficiency & Wave length of & Attenuation length\\
      $$ & (ns) &  (Photons/MeVee) & maximum emission (nm) & (cm) \\
      \hline
      EJ-262 & 2.1 & 8700 & 481 & 250\\
      EJ-240 & 285 & 6300 & 430 & 240\\
%      $\Delta$ t & 1.5 (FWHM) & \\
      \hline
    \end{tabular}
    \end{center}
%\end{ruledtabular}
\end{table*}

% \begin{figure}[!ht]
%        \centering
%          \includegraphics[width=8cm]{figures/PMT_UpDown_TimeDifference.png}          
%        \caption{(top panel) PCB board for Si preamps, (bottom left) the other side of the same PCB (vacuum side), (bottom right) PCB board covered with an aluminum plate for electrical noise reduction. 
%\cite{Glatz1986}. %
%        }\label{fig:Fig5}
%    \end{minipage}
%    \hfill{}
%\end{figure}

Another usage of the phoswich as a zero-degree detector is to monitor the beam rate with clear particle identification in an impure radioactive ion beam environment. 
This enables the determination of absolute elastic scattering cross sections with low intensity ($<$10$^{4-5}$ pps) radioactive ion beams. 
The experimental setup for this measurement is shown in Figure~\ref{fig:Fig4}. 
%Because of the short decay time of the EJ-262 (1 ns) and the long decay time of the EJ-240 (300 ns), 
%we can use the phoswich as a $\Delta E-E$ detector that can be used with a high intensity of incident particles (10$^{4-5}$ particles per second (pps)). 

The output pulses from the phoswich are processed with QDC implemented on the Mesytec MDPP-16 VME module. 
On this QDC, short and tail pulses are processed separately, 
and the short pulse's height ($\Delta E$) and integrated electric charges of the tail pulse ($Q$ or $E$) are provided (see \cite{MDPP-16_2022} and below and Figure~\ref{fig:Fig4} (d)) . 

\subsection{Elastic scattering monitor}

Monitoring the count rate of elastic scattering events helps deduce the beam rate, 
especially when the beam intensity is high in stable beam experiments ($>$10$^9$ pps) and a straightforward measurement with a Faraday cup is unavailable \cite{Ota2021}.  
BlueSTEAl houses a $\Delta E-E$ detector consisting of two Ortec Si surface barrier (diameter $\phi$=8 mm) detectors (thickness: 150 $\mu$m and 500 $\mu$m) to deduce the beam intensity from elastic scattering. 
The detectors are normally placed about 4.2$^{\prime\prime}$ away from the target position (see Fig. 1 c), d), and h)). 
The angular range which the $\Delta E-E$ detector covers is adjustable between  65--80 (mid angle) $\pm$2.5$^\circ$ from the target. 
When variations in the elastic scattering cross sections are large in this angular range, we cover the $\Delta E$ detector's surface with an aluminum collimator with a few mm $\phi$ hole at the center to narrow the angular range. 
By using a plate with an appropriate thickness which is a compromise between reducing the scattering at the edge of the hole and stopping the elastic particles, 
we usually observe a few counts / sec elastic events at 1 particle nA beam rate.  
This elastic-event rate is statistically enough to deduce the rate and spatial and temporal stability of the beam. 
It is important to use a $\Delta E-E$ detector to monitor the elastic scattering 
because e.g., in a $(d,p)$ experiment, $d$ is the elastic event while there are almost always $p$ from hydrogen contaminants in CD$_2$ targets \cite{Ota2021}, creating background events under elastic energy peaks. 

\begin{figure}[!ht]
        \centering
          \includegraphics[width=8cm]{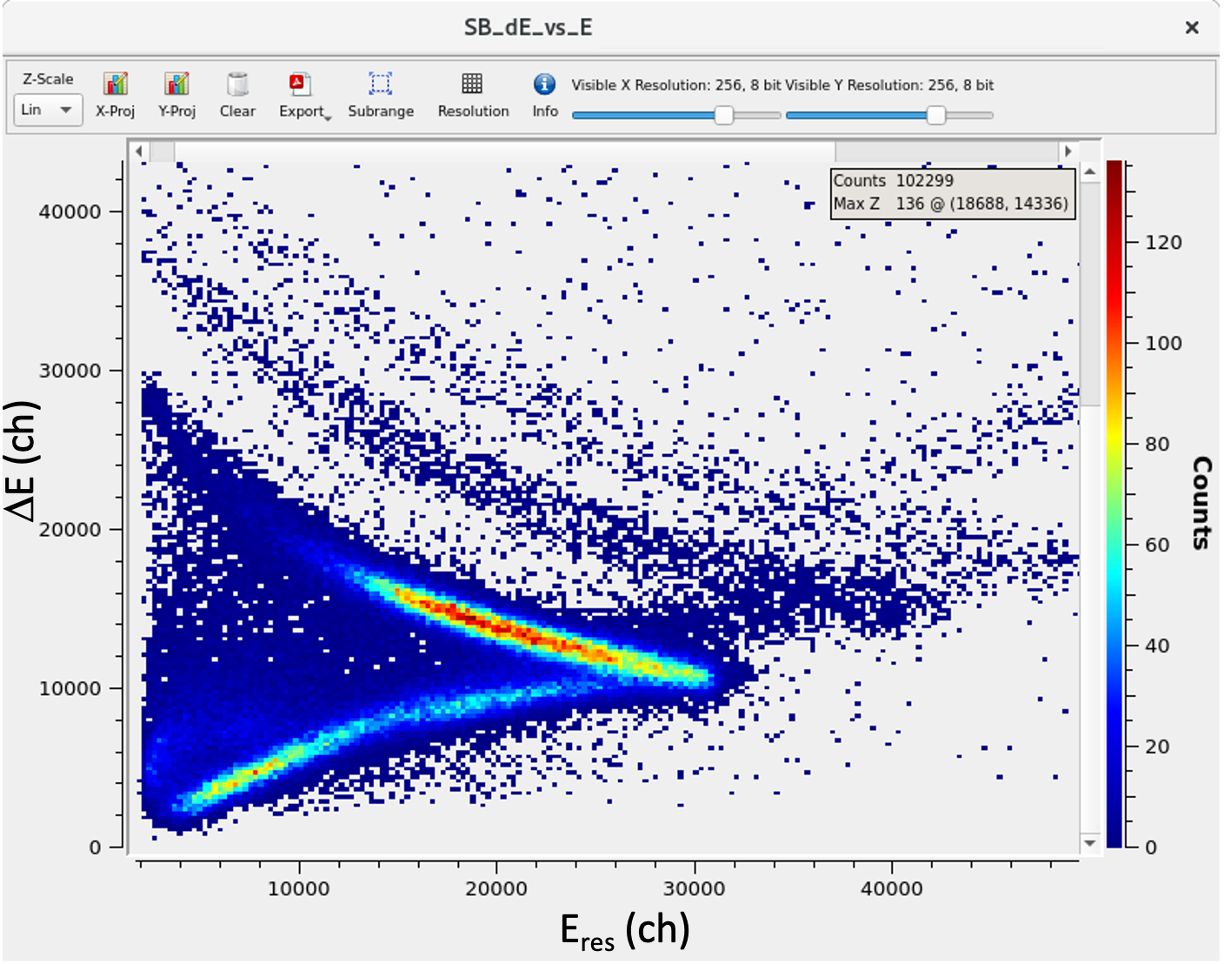}          
        \caption{Particle identification (protons, deuterons, and tritons) plot ($\Delta E-E_{res}$) by the Si surface barrier detectors monitored in real time on Mesytec DAQ (MVME). 
%\cite{Glatz1986}. %
        }\label{fig:Fig5}
%    \end{minipage}
    \hfill{}
\end{figure}

Since the detector outputs are read through Microdot connectors, 
the cables are adapted to Lemo cables (RG174) inside the BlueSTEAl chamber to readout through Lemo feedthroughs attached to one of the KF40 ports. 
Typical energy resolution achieved with a CAEN preamplifier and MDPP-16 SCP module (see below) was $<$ 30 keV FWHM for both of the detectors. 
Thus, the particle identification by these detectors is excellent as shown in Figure~\ref{fig:Fig5}.

\subsection{Diamond detectors}

 \begin{figure}[!ht]
        \centering
          \includegraphics[width=8cm]{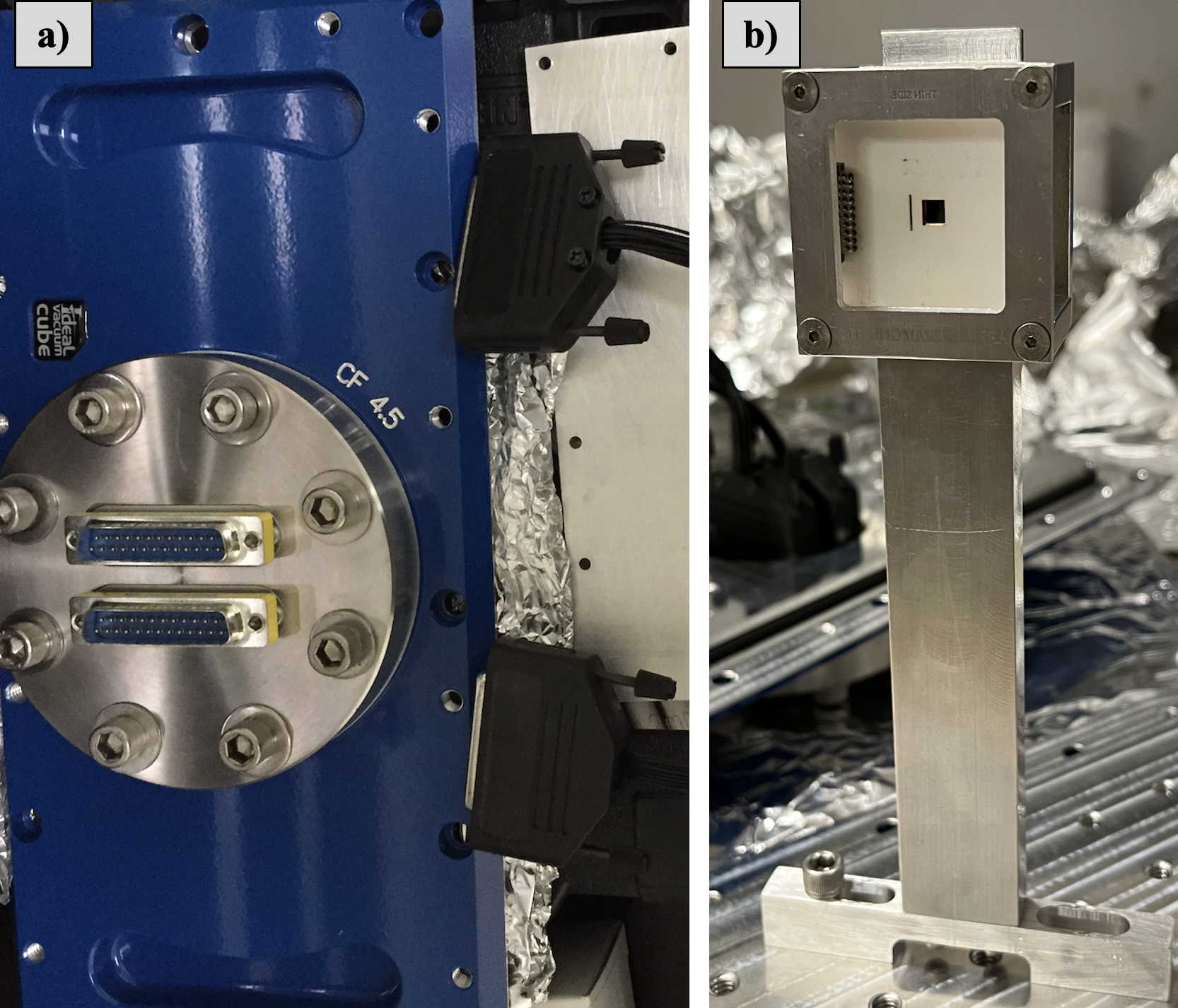}          
        \caption{a) CF4.5 feedthrough for the diamond detectors. Each Dsub25 connector delivers the signal from one diamond detector (16 ch). Inside the chamber (vacuum side), the signals are transported via the black cables in this photograph (Samtec SFSD-10-28-G-0600-S ribbon cables). b) The diamond detector attached to its detector holder. The signals are read out with Samtec TFM-110-01-L-D-RA connectors (see also Figure~\ref{fig:Fig1} c)).  
%\cite{Glatz1986}. %
        }\label{fig:Fig6}
%    \end{minipage}
    \hfill{}
\end{figure}

BlueSTEAl also houses a pair of thin (20 $\mu$m) and thick (500 $\mu$m) single-crystalline diamond (chemical vapor deposition (CVD)) detectors from Applied Diamond Inc. 
The dimension of both detectors is 4 mm$\times$ 4 mm (crystal itself) with a 38.1 mm$\times$38.1 mm PCB board, and each detector is segmented by 16 strips in either x or y direction. 
Thus, it is possible to identify the beam position with a resolution of 250 $\mu$m. 
The detectors are intended to be used to monitor the beam rate and observe the beam spot size more directly than with the elastic scattering monitor. 
Determining the beam position precisely helps improve the accuracy of the measured scattering angles, leading to improved angular distribution measurements, particularly at forward angles, as demonstrated in Section 4. 
Improving position resolution also leads to better energy resolution in reconstructed $E_x$ in a binary reaction from the measured energies with the Si detectors in inverse kinematics experiments.  
Although the detectors have not been tested with an actual beam experiment yet, we primarily expect to use the detectors for low intensity ($<$5$\times$10$^{3}$ pps) radioactive ion beam experiments such as elastic scattering measurements. 
Since Abbott et al. \cite{Abbott2022} report that the energy resolution rapidly degrades with heavy ion (7.5 MeV/u $^{78}$Kr) beams above $\sim$ 10$^3$ pps, it may be difficult to use for reactions with small cross sections ($<$ mb/sr), especially for heavy ($Z>$20) ion beams. 
Detailed responses to heavy ion beams of the same type of (but not segmented) diamond detectors have been reported, and the performances for our purposes are well verified in their study. 

 \begin{figure}[!ht]
        \centering
          \includegraphics[width=8cm]{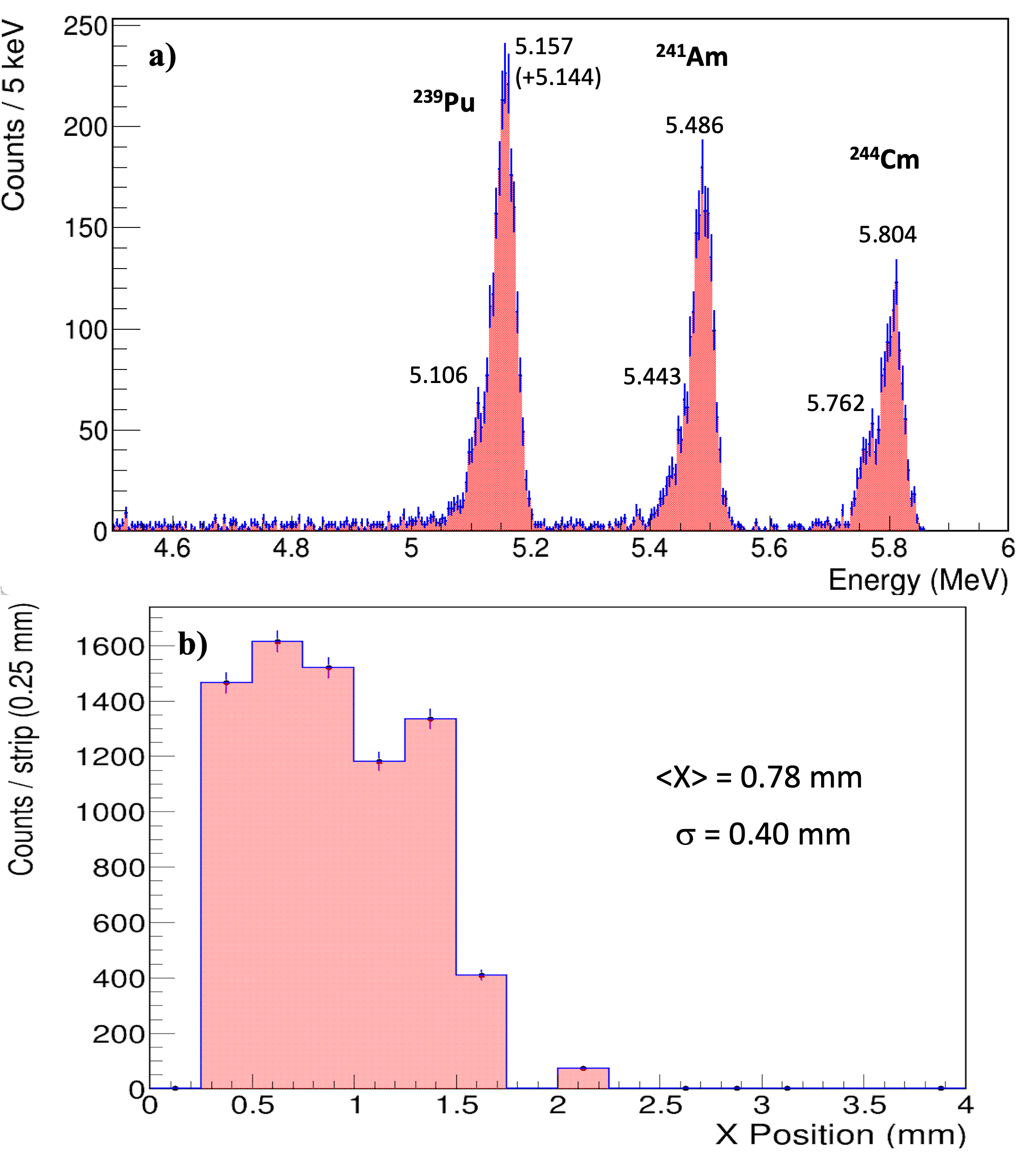}          
        \caption{a) $\alpha$ spectrum from the multi-nuclide source measured by the diamond $\Delta$ E detector, b) a distribution of $\alpha$ particles detected in each strip of the detector after passing the 1 mm $\phi$ slit (see text for detail). 
%\cite{Glatz1986}. %
        }\label{fig:Fig7}
%    \end{minipage}
    \hfill{}
\end{figure}

As a simple test of the detector's position resolution, 
we placed a collimator (a thin (1milli-inch)  aluminum foil) with a $\phi$=1 mm hole at the center between the $\alpha$ source attached to the target position and one (the thinner one) of the diamond detectors ($\sim$2.5 mm distance). 
The 16 outputs were read through the feedthrough in Figure~\ref{fig:Fig6}. 
Figure~\ref{fig:Fig7} shows an energy spectrum and position distribution measured by strip number (0.25 mm each). 
From the test, 
the energy resolution was obtained to be 40 keV FWHM 
and the position distribution ($\sim$1 mm) was reproduced well.

\subsection{Electronics and data acquisition system}

Mesytec high-voltage supply modules (MHV4s) were used to bias all detectors described above except for the phoswich (which is typically biased with in-house precision high voltage power supplies such as 375X HV modules, Bertan Associates, Inc.). 
Mesytec preamps (MPR-16 or MPR-32) were used for the S2 Si detectors and diamond detectors, while CAEN charge sensitive preamp (MOD A1422 8 ch 90 mV/MeV) was used for the ORTEC Si detectors (elastic scattering monitor). 
Pulses from the preamp / PMT output are fed to a 200 MS digitizer VME SCP/QDC module (MDPP-16 or MDPP32, Mesytec). 
These pulses are processed to a triangle-shaped pulse after initial noise filtering on MDPP SCP (Sensitive Charge Amplifier) modules for detectors except for the phoswich, whose PMT output pulses are processed by a MDPP QDC module. 

\begin{figure*}[!ht]
        \centering
          \includegraphics[width=16cm]{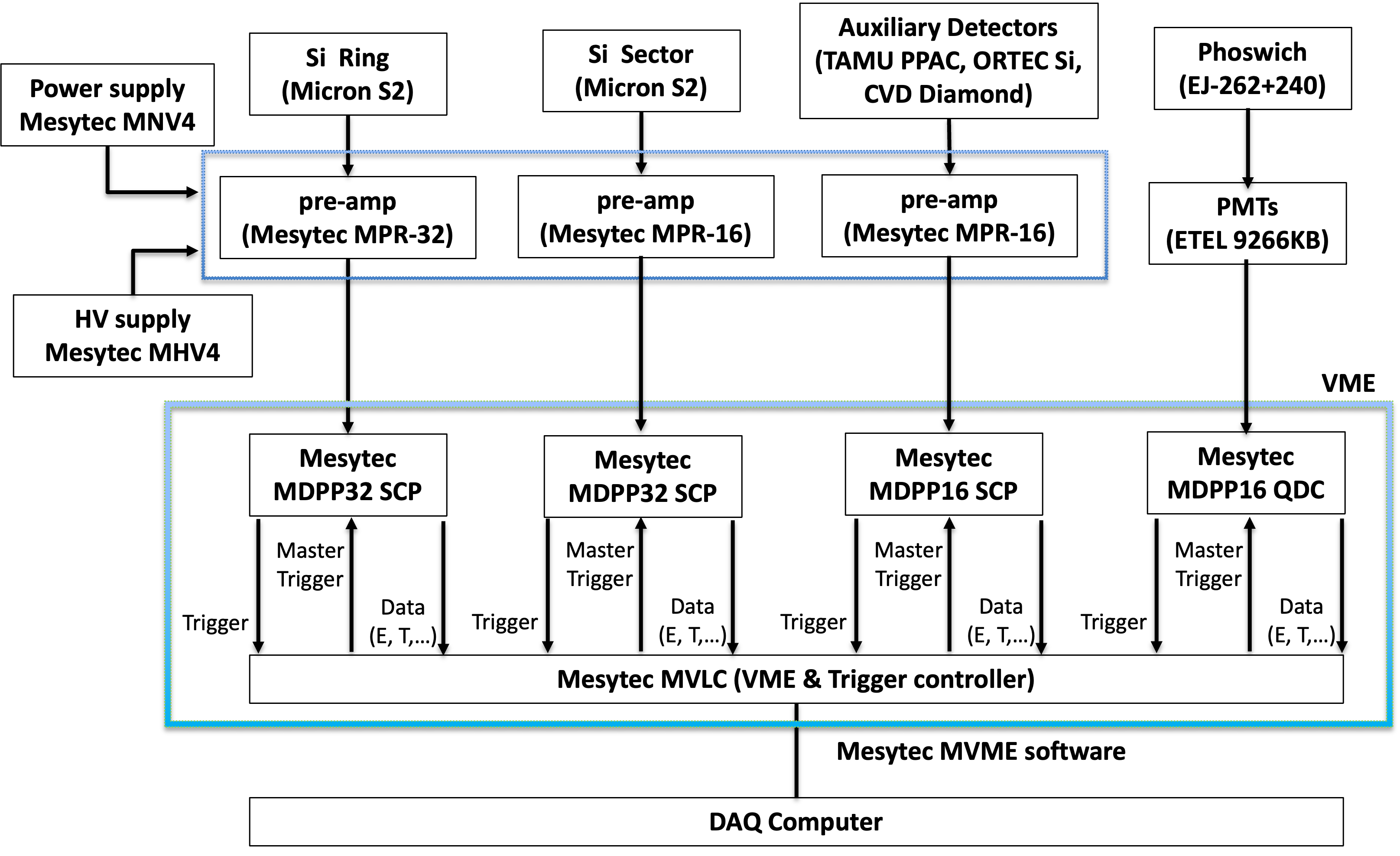}          
        \caption{Diagram of the BlueSTEAl electronics. 
%\cite{Glatz1986}. %
        }\label{fig:Fig8}
%    \end{minipage}
    \hfill{}
\end{figure*}

Both SCP and QDC modules provide both energy and timing information from every single input channel. 
The energy and timing are evaluated with a high resolution ADC (16 bit) and time-to-digital converter (TDC (16 bit, 24 ps resolution (finest))), respectively.
The events are self-triggered on each input channel, and 
the data are recorded if the event timing measured with internal CFDs (Constant Fraction Discriminators) falls on the trigger window opened internally (by the earliest-arrived event on the module) or by external trigger. 
Various pulse shaping parameters including gain, shaping time, and inter-module trigger settings are controlled through a text editor on the DAQ software, MVME (version 1.4.7 was used on our Linux platform DAQ computer server). 
All the information of energies, timings, and a trigger from each SCP/QDC module are collected in the VME controller module MVLC, which is accessed and controlled with MVME via internet or USB connection. 
The triggers processed (e.g., making triggers from different modules ORed) by MVLC are used as a master trigger, which is fed to the individual modules as an external trigger. 
The input trigger timing is recorded with a timestamp extracted from a clock on the VME module (16 MHz) on each module. 
Based on the timing information, data across the VME modules are recorded as an event. 
Figure~\ref{fig:Fig8} summarizes the electronics and DAQ configuration of the BlueSTEAl detector system. 

While the MVME DAQ software displays energy and timing spectra from each channel real-time (see e.g., Figure~\ref{fig:Fig5}), 
the data simultaneously recorded in a hard-drive on the DAQ computer (Linux platform) server are available in the CERN ROOT format or a custom format developed by Mesytec. 
The DAQ dead time per trigger is typically below a few $\mu$s (dominated by the trigger window set at single event mode \cite{MDPP-16_2022}) and thus, in principle, 
$\sim$10$^5$ events / second can be recorded without critical dead time issues. 
In a simple test we performed with an external periodic linear pulse generator (note $\it{not}$ random pulse; the input pulses were sent through one of the MPR-16 pulser input), 
we obtained the livetime of our DAQ to be 100\% up to 25 kHz input pulses. 
Therefore, it would be practical to use our detector system below 5$\times$10$^4$ trigger rate. 
%In an actual experiment described below, we performed the $\^{21}$Ne$(p,t)$ experiment at $\sim$ 
%To verify the deadtime, we performed a simple test of the livetime with an external periodic linear pulse generator.  
%The DAQ livetime was measured as a function of the pulser rate (a ratio of the trigger rates to the input pulse rates). 
%The input pulses were sent through one of the MPR-16 pulser input. 
%Therefore, only one of the MDPP-16 modules were triggered and processed the pulses. 
%The livetime was obtained to be 100\% up to 25 kHz input pulses, 66\% at 50 kHz, and rapidly dropped above the input rate. 
%Thus, it would be practical to use our detector system below 5$\times$10$^4$ trigger rate. 
It should be noted that this becomes a limiting factor primarily when measuring the beam count rates for absolute elastic scattering cross section measurements, 
where beams are used as a trigger to create a gate window. 
In many other cases such as binary reactions in inverse kinematics experiments, 
the events are typically triggered by light ejecta detected with Si detectors, whose rate should be much smaller. 
In these cases, the limiting factor on increasing the beam rate is rather the focal-plane detectors' radiation resilience and pulse pile up. 
Further specifications of the DAQ system are available on the Mesytec website \cite{MVME2020}. 

\section{Si detector array efficiency}
\subsection{Stable beam experiments}

\begin{figure}[!ht]
        \centering
          \includegraphics[width=8cm]{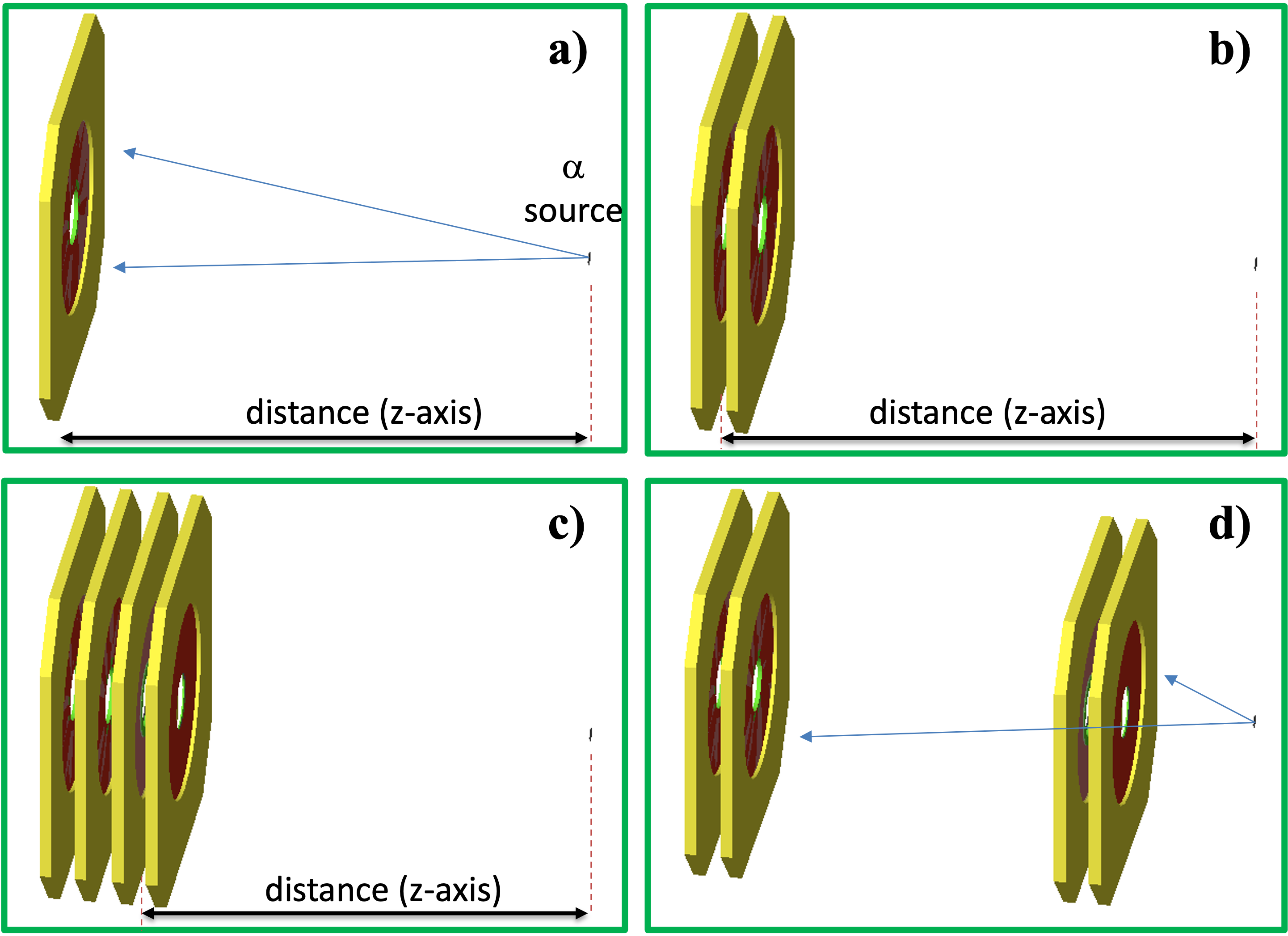}          
        \caption{Typical Si detector configurations; a) a single detector setup with the BlueSTEAl, b) a double detector setup, c) a quadruple detector setup (a stack of four Si detectors), and d) a pair of double detector setup. The Si detector holders are not included in the drawing. An $\alpha$ source ($\phi$$\sim$5 mm) assumed in the simulations is shown together. 
%\cite{Glatz1986}. %
        }\label{fig:Fig9a}
%    \end{minipage}
    \hfill{}
\end{figure}

Owing to the four S2 Si detectors each of which is highly segmented, 
different configurations of Si arrays can be built (Figure~\ref{fig:Fig9a}). 
While our most commonly used setup so far is to use the four detectors as a thick (5 mm) bundle of the detectors ($\Delta E-\Delta E-\Delta E-E$) (Figure~\ref{fig:Fig9a} c)) or 
make two pairs of $\Delta E-E$ detectors placed separately (Figure~\ref{fig:Fig9a} d)), 
these configurations lose detection efficiency by requiring coincidence among the bundled/paired detectors. 
In this section, the efficiency is evaluated for these configurations 
using Monte Carlo simulations. 

The simulations were made with the Geant4 framework on the nptool platform \cite{Matta2016}. 
In the simulations, an isotropic $\alpha$-source was virtually set at the BlueSTEAl target position ($x,y,z$=(0,0,0); 10$^7$ $\alpha$ particles / simulation), 
and then the number of $\alpha$ particles detected by each detector set (in coincidence in cases where multiple detectors are used)
was divided by the number of $\alpha$ particles emitted at a given angle (known from the simulation). 
The $\alpha$-particle energy was adjusted to be stopped in the detector placed at the most downstream position (i.e., residual energy detector). 
Note the ratio of the number of events detected by a given detector ring to the total number of emitted $\alpha$ particles (10$^7$ events) represents a solid angle subtended by the ring by multiplying 4$\pi$, when the $\alpha$ source has no spatial spread. 
The solid angles obtained in this method were later used for elastic cross section measurements. 
Meanwhile, the goal of this exercise is to obtain the detection efficiency of the Si detector array
when imposing realistic beam conditions (beam position, particularly spatial distribution). 
Note in the following simulations, the first ring (the one with a single strip) is not included. 
This is because, in many situations, we ignore the first ring due to dominant stray beam (beam halo) events. 
Indeed, the first ring often severely suffers from electric charge sharing from events that hit the inner, electrically insulated rings (see above; very forward angles) in the high-flux beam environment. 

\begin{figure*}[hbt!]
        \centering
          \includegraphics[width=15cm]{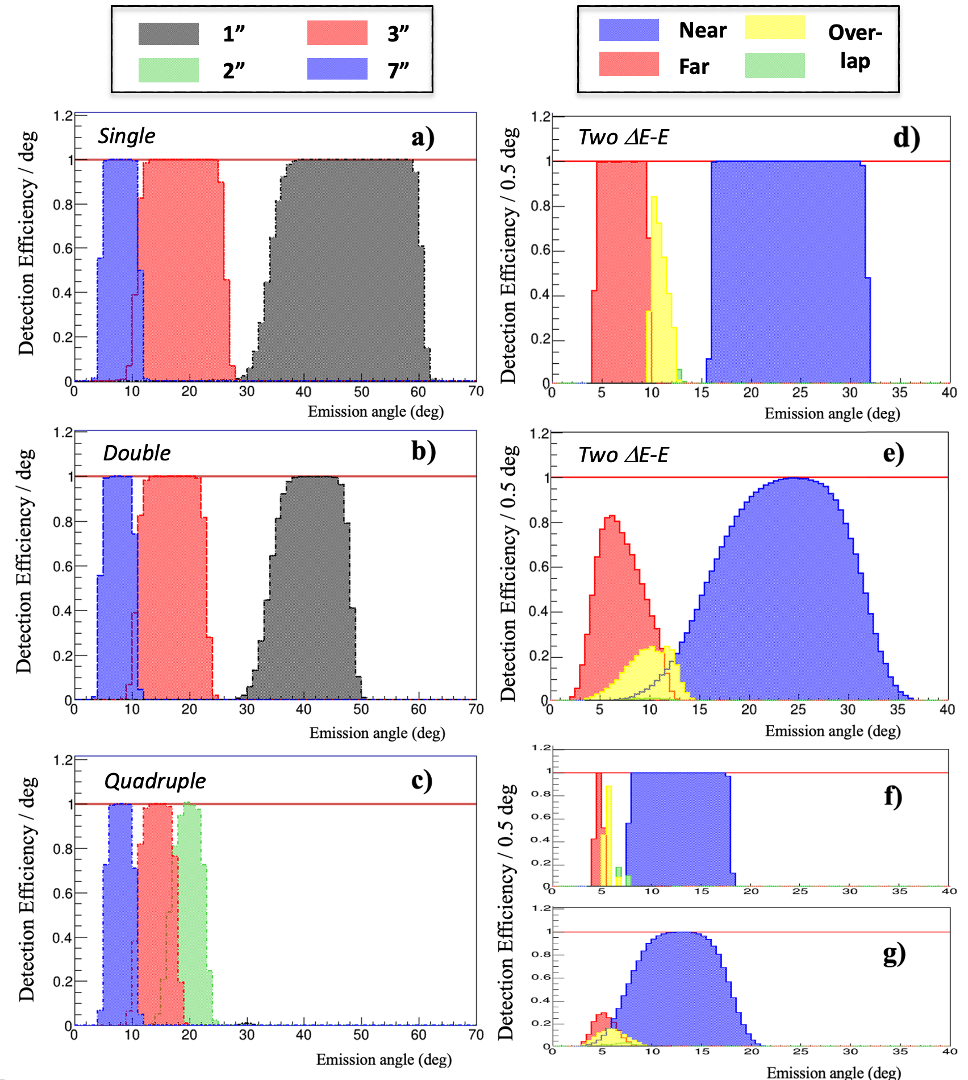}          
        \caption{Evaluated angular coverage and detection efficiency by BlueSTEAl S2 detectors from Monte Carlo simulation; a) A single detector was located at 1$^{\prime\prime}$ (gray), 3$^{\prime\prime}$ (red), and 7$^{\prime\prime}$ (blue) downstream from the target position, 
        b) A double ($\Delta E-E$) detector placed at 1$^{\prime\prime}$ (gray), 3$^{\prime\prime}$ (red), and 7$^{\prime\prime}$ (blue) from the target, c) A quadruple ($\Delta E-\Delta E-\Delta E-E$) detector placed at 1$^{\prime\prime}$ (gray), 2$^{\prime\prime}$ (green), 3$^{\prime\prime}$ (red), and 6$^{\prime\prime}$ (blue) from the target. Note there is almost no efficiency at 1$^{\prime\prime}$ in this setup ($\sim$30$^{\circ}$), 
        d) A pair of the double detectors placed at the near side (2$^{\prime\prime}$ (blue)) and far side (7$^{\prime\prime}$ (red)) from the target, respectively, assuming no beam spread. The angles overlapped between the two $\Delta E-E$ detectors are shown in yellow (the inner area than the active rings of the 2nd Si detector are hit; therefore not detected in the real experiment) and green (the active rings of the 2nd Si detector are hit).
        e) the same as d) but with $\phi$=3 mm beam size (rms), 
        f) and g) A pair of the double detectors placed at 4$^{\prime\prime}$ and 7$^{\prime\prime}$ from the target assuming $\phi$=0 and 3 mm beam size (rms), respectively. The color coding is the same as d) and e). 
%\cite{Glatz1986}. %
        }\label{fig:Fig9}
%    \end{minipage}
    \hfill{}
\end{figure*}

First, we placed only one single detector in normal to the beam axis ($z$-axis) and shifted it to various possible positions along the axis, 
from 1$^{\prime\prime}$ to 7$^{\prime\prime}$ downstream from the target position (Figure~\ref{fig:Fig9a} a)). 
The beam position was centered as $x,y$=(0,0), where the centers of both target and all S2 detectors reside, with 1 mm spread in $x$ and $y$ directions (rms) with a normal distribution. 
Note that this spread is larger than normal stable beam's cases (Figure~\ref{fig:Fig3}). 
Figure~\ref{fig:Fig9} a) shows the detection efficiency at 1$^{\prime\prime}$, 3$^{\prime\prime}$, and 7$^{\prime\prime}$ downstream from the target. 
Note $x$-axis denotes the angles at which $\alpha$ particles are emitted in the laboratory system (equivalent to scattering angles in case of beam experiments), and not the angles faced by the detector rings with regard to $x,y,z$=(0,0,0). The two angles are similar but slightly different when the source/beam has spatial spread. 
In the figure, because of the assumed $\alpha$ source's spatial spread, the efficiency gradually (not discretely) reduces to 0 at the lower and upper edge of the detector. 
At the 7$^{\prime\prime}$ distance, 4--11$^{\circ}$ is covered, while 35--60$^{\circ}$ is covered at 1$^{\prime\prime}$ distance.

In the second simulation, two detectors are paired and attached to the detector holder (Figure~\ref{fig:Fig9a} b)). 
A mid-plane between the two detectors along the $z$-axis was placed somewhere 
between 1$^{\prime\prime}$ and 7$^{\prime\prime}$ from the target and moved by 1$^{\prime\prime}$ in each simulation. 
Taking into account the detector holder's thickness, each detector is positioned $\pm\sim$0.25$^{\prime\prime}$ upstream and downstream from the mid-plane along $z$-axis. 
Figure~\ref{fig:Fig9} b) shows the detector efficiency of the $\Delta E-E$ detector at various distances from the target. 
In comparison with the single detector setup, 
the efficiency only moderately reduces at 7$^{\prime\prime}$ distance, whereas the efficiencies at 1$^{\prime\prime}$ and 3$^{\prime\prime}$ significantly reduce at higher angles. 
The reduction primarily occurs because the large angle events detected at the front ($\Delta$E) Si detector go beyond the rear ($E$) detector's upper edge (the 23rd ring) and do not hit the rear detector. 
The efficiency reduction is more significant when the four Si detectors are bundled (Figure~\ref{fig:Fig9a} c)). 
In this setup, each of two detectors are attached to one detector holder, 
and the two holders were placed 1$^{\prime\prime}$ distant along $z$-axis from each other.  
The distance from the $\alpha$ source was defined by the midplane position of the front two detectors. 
As shown in Figure~\ref{fig:Fig9} c), the efficiency at 1$^{\prime\prime}$ distance reduces to almost zero (see $\sim$30$^{\circ}$). 
The efficiency at 3$^{\prime\prime}$ distance is also significantly reduced, but still some events can be detected at forward angles (11--18$^{\circ}$).
The efficiency at 6$^{\prime\prime}$ is not significantly reduced even compared to the single detector setup. 
Thus, the configuration at 6$^{\prime\prime}$ from the target is the most beneficial when one needs to use the four detectors as a bundle. 
Indeed, the $^{21}$Ne($p,t$) experiments discussed later used this setup. 
A summary of the simulation results for all conditions are shown in Table~\ref{tab:table2}.

\subsection{Radioactive ion beam experiments}

When radioactive ion beams are used for an experiment, 
the common problems are low beam intensity and relatively large beam spread. 
Assuming such experiments, 
the simulations were made by placing an $\alpha$ source at the beam center ($x,y$=0,0) with 3 mm spread (rms) in both $x$ and $y$ directions. 
The spatial spread is based on the actual beam size observed in the $^{11}$Be experiment as described in Section 4. 
The two paired-detectors ($\Delta E-E$) were set 2$^{\prime\prime}$ and 7$^{\prime\prime}$ downstream from the target, respectively, to cover broad angles in one single experiment (Figure~\ref{fig:Fig9a} d)). 
Ten million $\alpha$ particles were generated and emitted isotropically in each simulation. 

In the simulation shown in Figure~\ref{fig:Fig9} d), the beam spread was reduced to 0 mm for clarity. 
In this setting, while the angular coverage by the downstream $\Delta E-E$ detector (red) is slightly reduced ($\theta$$\sim$12$^\circ$; see Figure~\ref{fig:Fig9} b) (blue)) by being shadows of the inner ring areas of the 2nd Si detector (the $E$ detector in the upstream detectors) (yellow), the two $\Delta E-E$ (red and blue) cover broad scattering angles (4--10$^\circ$ and 17--32$^{\circ}$) independently without overlap. 
%Interestingly, some small angle gaps (at $\sim$7$^{\circ}$) between the two detector sets are covered by the 2nd and 3rd Si detectors (the $E$ detector in the upstream detectors and the $\Delta$E detector in the downstream detectors). 
For each detector set, the efficiency discretely reduces to 0 at either the lower or upper end of angles because the $\alpha$ source has no spatial spread. 
In Figure~\ref{fig:Fig9} e), the beam spread was set to 3 mm (rms) in both $x$ and $y$ directions. 
The angular coverage by each of the $\Delta E-E$ detector sets broaden and some overlapping events between the different pairs of detectors (events emitted at the same angle but detected by different detector sets) are found. 
Such overlapping events are found between the 2nd and 3rd detectors (yellow) as well, leading to an efficiency loss of the downstream $\Delta E-E$ detector set. 
Indeed, the efficiency loss of the detector set by being shadowed by the 2nd detector amounts to 50\% at large angles ($\theta$$\gtrsim$10$^\circ$) in this setting, which was confirmed by simulating with and without the upstream detector set. 

Figure~\ref{fig:Fig9} f) shows the case where the first detector set was placed 4$^{\prime\prime}$ downstream from the target instead of 2$^{\prime\prime}$ (the other set is kept 7$^{\prime\prime}$ from the target). The beam size was set to 0 for clarity. 
In this setting, the angles covered by the downstream $\Delta E-E$ detector are significantly reduced by being the shadow of the upstream detector set, and only $\theta$$\sim$4--6$^\circ$ are detected. 
When the beam size was set to 3 mm (rms) in both $x$ and $y$ directions, the efficiency of the detector set further significantly reduced as shown in Figure~\ref{fig:Fig9} g). 
Therefore, for radioactive ion beam experiments, the 2$^{\prime\prime}$--7$^{\prime\prime}$ distance setting is simpler and more efficient, and such a setup was used in the actual $^{11}$Be experiment below. 
However, the efficiency loss still needs to be carefully estimated and corrected. 
Therefore, in the $^{11}$Be experiment, we performed measurements with only one $\Delta E-E$ to estimate the efficiency loss as well. 

%It is also important to note that the angles shown in the figures are the angles the detectors are facing toward the beam center (apparent angle). %and used for data analysis. 

In summary, it should be noted that the detection efficiency obtained in this section is based on the emission/scattering angles of the particles. 
In the actual experiment without accurate information of the beam position/spread, 
the detection efficiency obtained based on the angles faced by the detectors (which is geometrically obtained by assuming a point beam source as in Figure~\ref{fig:Fig9} d and f)) should be slightly different from those we obtained in this section (Figure~\ref{fig:Fig9} e and g)). 
For example, while we believe events that fall on the red area in Figure~\ref{fig:Fig9} d) ($\theta$$\sim$4--10$^\circ$) are detected, events in the red area in Figure~\ref{fig:Fig9} e)  ($\theta$$\sim$2--12$^\circ$) are detected in reality. 
Such differences can impact the accuracy of determining the cross sections, especially at forward angles where cross sections steeply vary by a small angular difference, as discussed in the following section (the $^{11}$Be experiment). 
Moreover, the sensitivity of measured cross sections to angular distribution is degraded. 
Thus, implementing a beam tracking device such as the diamond detectors and PPACs with the BlueSteAl will be an important task in the future. 

\begin{table*}
%        \centering
  \begin{center}
    \caption{BlueSTEAl Annular S2 Si array angular coverage and detection efficiency estimated by Monte Carlo simulations, assuming the $\alpha$ particle source was centered and has a 1 mm spatial spread ($x,y$, rms) on the target position. The geometrical efficiency was obtained by dividing the total number of the $\alpha$ particles detected with detectors (in coincidence, except for the single detector setup) by the number of $\alpha$ particles emitted isotropically from the source (10$^7$ particles). By multiplying it with 4$\pi$, the solid angle coverage is obtained (not exact due to the assumed beam spread). The $\theta$ coverage was defined by a range between the mid-angles of the paired innermost ring (2nd ring) and the effective outermost ring. The first single strip (unpaired) ring (1st ring) was not considered (see text). The effective outermost ring was defined by the outermost rings hit in the first ($\Delta$E) detector. They vary when requiring coincidence detection with following detectors. The average $\theta$ resolution ($\delta\theta$) is defined by dividing the $\theta$ coverage with the number of effective rings (from the 2nd ring to the effective outermost rings).\\} 
    
    \label{tab:table2}
%    \begin{ruledtabular}
    \begin{tabular}{cccccc} % <-- Alignments: 1st column left, 2nd middle and 3rd right, with vertical lines in between
          \hline
      Configuration & Distance from  & Geometrical & $\theta$ coverage ($^{\circ}$) & Effective & $\delta \theta$ ($^{\circ}$)\\
       &  target (mm) & efficiency (\%) &  & outermost rings & \\      
      \hline
      Single & 25.4 & 17.08 & 35.07--60.32 & 23 & 1.20 \\
      & 50.8 & 8.45 & 17.20--37.45 & 23 & 0.96 \\
      & 76.2 & 4.34 & 11.24--26.10 & 23 & 0.71 \\
      & 101.6 & 2.55 & 8.32--19.81 & 23 & 0.55 \\
      & 126.0 & 1.66 & 6.58--15.89 & 23 & 0.44 \\
      & 151.4 & 1.15 & 5.44--13.25 & 23 & 0.37 \\  
      & 176.8 & 0.85 & 4.66--11.35 & 23 & 0.32 \\                              
      Double & 25.4 & 8.28 &  35.07--47.79 & 10,11 & 1.50 \\
      & 50.8 & 5.41 & 17.20--31.36 & 16,17 & 0.98 \\
      & 76.2 & 3.13 & 11.24--22.82 & 18,19 & 0.70 \\
      & 101.6 & 1.98 & 8.32--17.80 & 19,20 & 0.53 \\
      & 126.0 & 1.35 & 6.58--14.57 & 20,21 & 0.43 \\
      & 151.4 & 0.97 & 5.44--12.30 & 21,22& 0.35 \\
      & 176.8 & 0.73 & 4.65--10.66 & 21,22 & 0.31 \\      
      Quadruple & 25.4 & 0.02 & 30.11--30.28 & 2 & -- \\
      & 50.8 & 1.91 & 17.20--22.56 & 7,8 & 0.97 \\
      & 76.2 & 1.56 & 11.23--17.65 & 11,12 & 0.68 \\
      & 101.6 & 1.15 & 8.31--14.44 & 13,14 & 0.53 \\
      & 126.0 & 0.87 & 6.60--12.21 & 15,16 & 0.42 \\  
      & 151.4 & 0.67 & 5.46--10.59 & 15,16 & 0.38 \\                                    
%      $\Delta$ t & 1.5 (FWHM) & \\
      \hline
    \end{tabular}
    \end{center}
%\end{ruledtabular}
\end{table*}

\section{Commissioning}
\subsection{$^{21}$Ne($p,t$)$^{19}$Ne reaction}

 \begin{figure}[!ht]
        \centering
          \includegraphics[width=8cm]{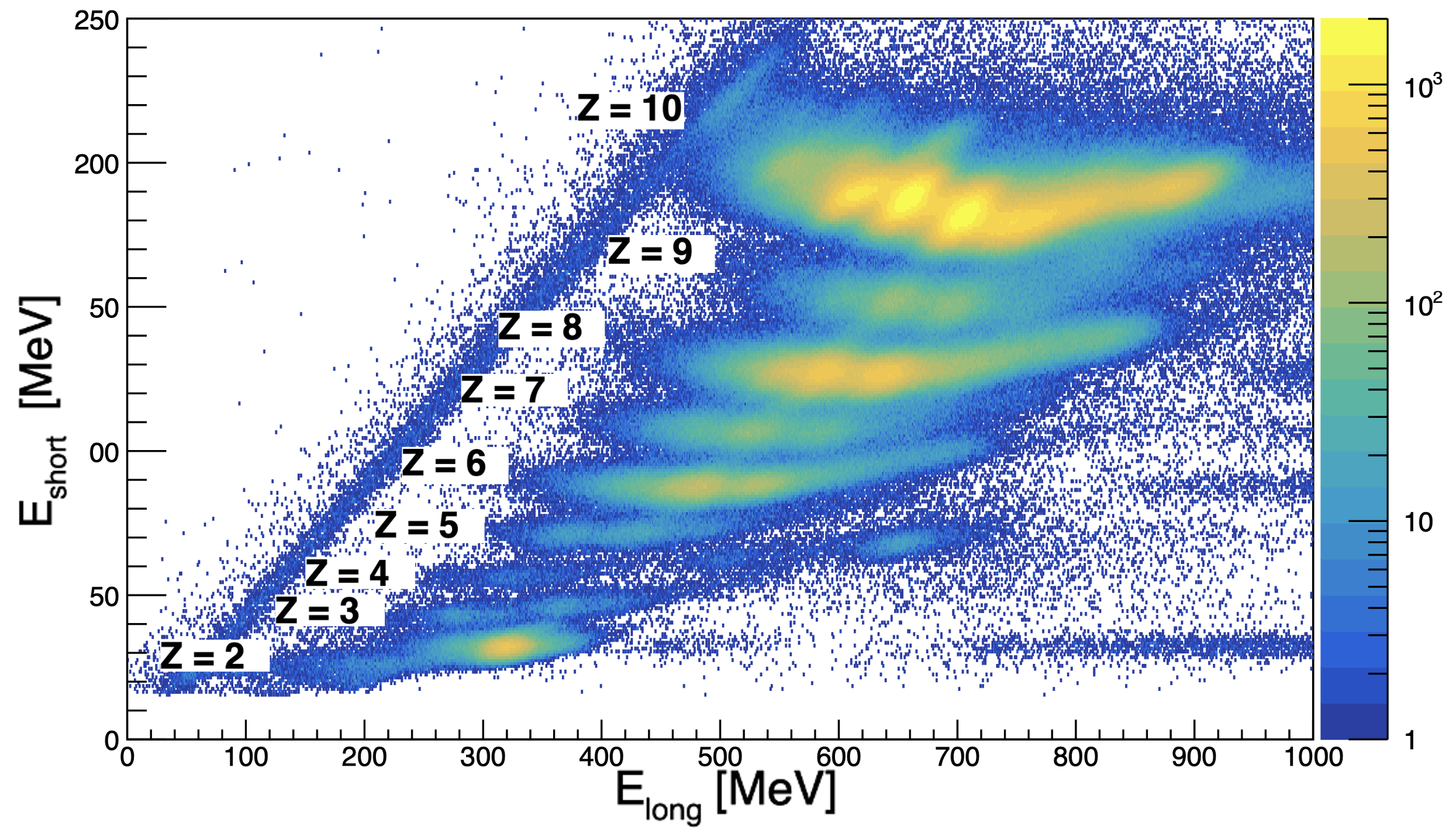}          
        \caption{Particle identification plot from the $^{21}$Ne$(p,t)$ experiment with the phoswich detector. 
%\cite{Glatz1986}. %
        }\label{fig:Fig10}
%    \end{minipage}
    \hfill{}
\end{figure}

As a first experiment with BlueSTEAl, 
we performed an experiment studying the $^{21}$Ne($p,t$)$^{19}$Ne reaction using the K500 cyclotron at TAMU cyclotron institute. 
The goal of the experiment is to determine the $\alpha$ and $\gamma$-ray decay branching ratios for resonances in the $^{15}$O($\alpha,\gamma$) reaction, which plays an important role for nucleosynthesis in binary stars (classical novae and Type-I X-ray bursts) \cite{Davids2011}.

A 40 MeV/u $^{21}$Ne beam impinged on a CH$_2$ target (1 mg/cm$^2$) at 10 particle nA.  
Because of the negative Q-value (-15.14467 MeV \cite{NNDC2023}), 
the reaction needs a high energy beam and all of the light ejecta are scattered at forward angles in the laboratory system. 
Therefore, a stack of our four Si detectors were placed inside the BlueSTEAl chamber to cover 10--30$^\circ$ in the Center-of-Mass (CM) system. 
The magnetic field in the MDM spectrometer was set to a rigidity $(B\rho)$ to accept both $^{19}$Ne and $^{15}$O recoils produced from the decay of excited $^{19}$Ne states produced in the ($p,t$) reaction. 

Figure~\ref{fig:Fig10} shows the obtained elemental ($Z$) identification plot from the fast (short) and slow (long) outputs in the phoswich detector. 
To reduce the sensitivity to the hit position, the outputs from the left and right ends of the phoswich were combined as $E=(E_{left}\times E_{right})^{1/2}$ in both fast and slow outputs. 
Ne and O are clearly identified from the plot, and isotopic identification based on the position information from the double PPAC is currently under analysis.

\subsection{$^{11}$Be elastic scattering on $^{12}$C}

Elastic scattering cross sections carry rich information to constrain optical potentials (e.g., \cite{Lapoux2008}) which are an essential piece of information to calculate scattering or reaction cross sections using theories such as Distorted Wave Born Approximation (DWBA). 
Not only that, precise measurements of angular differential elastic scattering cross sections which are typically order-of-magnitude larger than reaction cross sections, could lead to constraining various nuclear properties of halo nuclei with the limited statistics resulting from low-intensity radioactive ion beams \cite{Johnson1997}. 
Combined with one neutron breakup cross sections which are relatively large as well in a neutron halo system, 
high quality data of elastic and breakup angular differential cross sections are expected to be a new tool to diagnose halo properties \cite{Capel2011}. 

For this purpose, we used BlueSTEAl to measure elastic scattering cross sections in the $^{11}$Be+$^{12}$C system. 
A 22 MeV/u $^{11}$Be secondary beam was produced with a 30 MeV/u $^{13}$C primary beam on a Be target with the MARS spectrometer \cite{Tribble1989} using the K500 cyclotron  at TAMU cyclotron institute. 
The secondary $^{11}$Be beam intensity was $\sim$8$\times$10$^3$ pps (with $\sim$80\% purity; $^{8}$Li and light ions account for the remaining $\sim$10\% and $\sim$10\%, respectively) and the beam spot size was about 2.5--3 mm (rms). 
Two sets of Si $\Delta E-E$ detectors are placed 2$^{\prime\prime}$ and 7$^{\prime\prime}$ downstream from a 17 mg/cm$^2$ C target (0.5 mm + 1.5 mm for the upstream Si array to cover 17--31$^\circ$; 1.5 mm $\times$2 for the downstream array to cover 5--10$^\circ$ lab angles, respectively (Config. 1)).
Later, during the same experiment, only one $\Delta E-E$ detector (1.5 mm$\times$2) was placed 4$^{\prime\prime}$ downstream from the target (Config. 2) to cover 8--18$^{\circ}$.  
Because of the overlapping angles in these two configurations (Config. 1 and 2), the measured cross sections can be double checked. 

\begin{figure}[!ht]
        \centering
          \includegraphics[width=8cm]{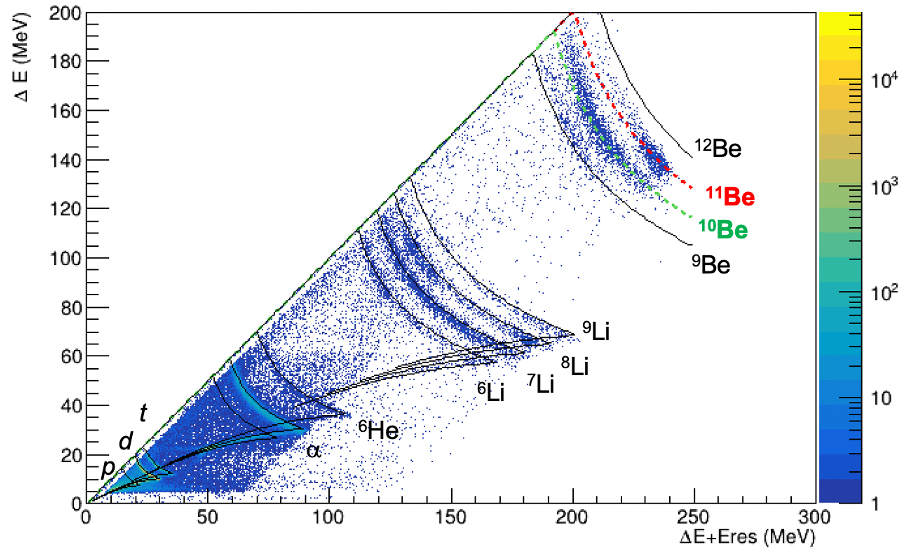}          
        \caption{Particle identification plot from the $^{11}$Be+$^{12}$C experiment with the annular S2 Si $\Delta E-E$ detectors. 
%\cite{Glatz1986}. %
        }\label{fig:Fig11}
%    \end{minipage}
    \hfill{}
\end{figure}

\begin{figure}[!ht]
        \centering
          \includegraphics[width=8cm]{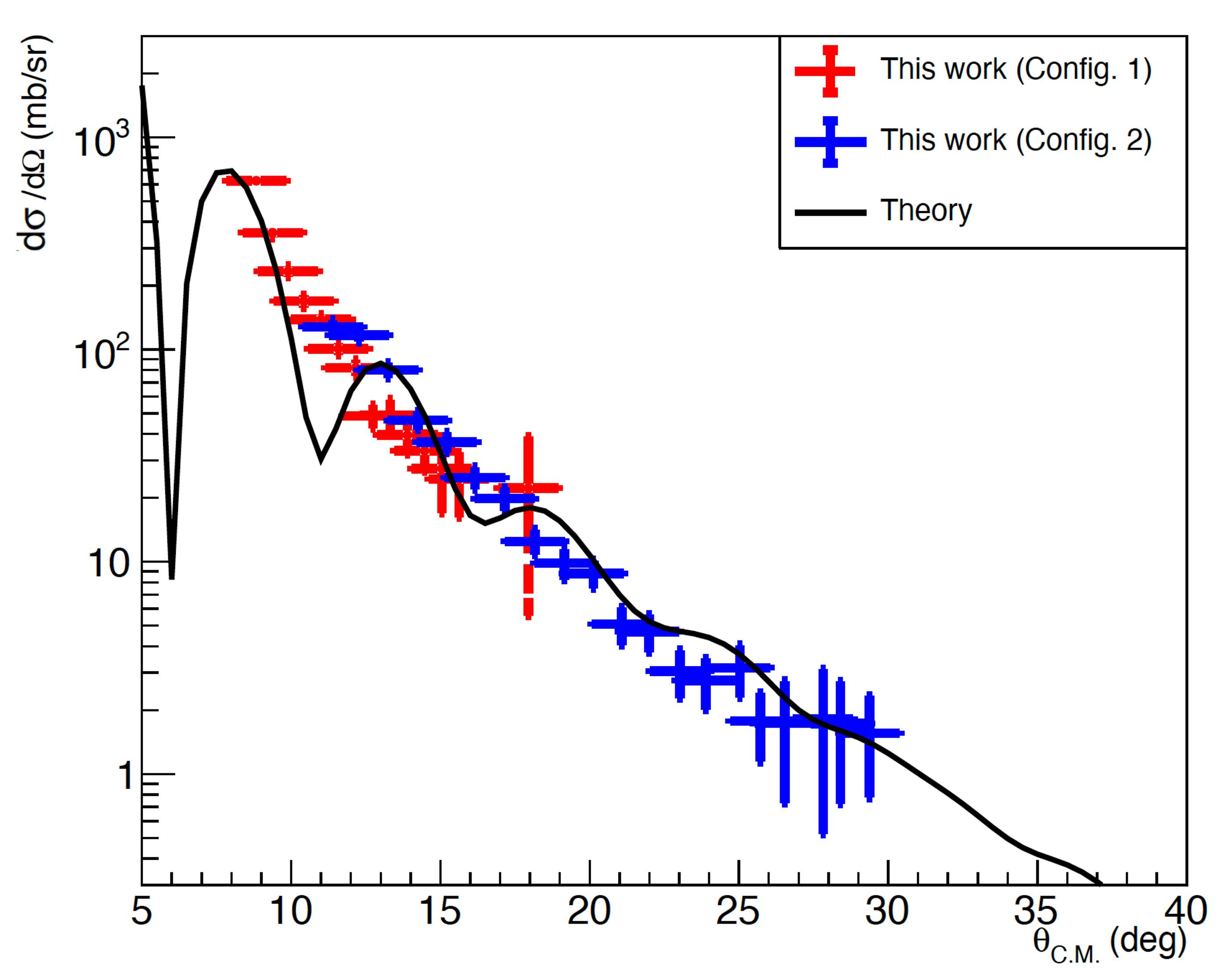}          
        \caption{Preliminary results of angular differential elastic scattering cross sections from the $^{11}$Be+$^{12}$C experiment. 
%\cite{Glatz1986}. %
        }\label{fig:Fig12}
%    \end{minipage}
    \hfill{}
\end{figure}

Figure~\ref{fig:Fig11} shows the $\Delta E-E$ plot from the downstream Si detector array of the Config. 1. 
All isotopes up to $^{11}$Be are clearly identifiable from the plot. 
Figure~\ref{fig:Fig12} shows preliminary results of the absolute elastic scattering cross sections (in the CM frame) obtained from the experiment 
after correcting for the beam position (which was found slightly off-centered). 
The error bars in the cross sections are statistical uncertainties and the errors in the angles are evaluated from Monte Carlo simulations assuming a realistic beam position, spatial and angular spread. 
The data at large angles from the Config. 1 is not shown. 
Note $^{11}$Be has a bound excited state and the state is not distinguished from the ground state in our detector system. 
Nevertheless, since the inelastic scattering cross sections are negligible compared to the elastic cross section, 
the measured cross sections represent the elastic cross sections. 
It should also be noted that correction of the DAQ dead time (observed $\lesssim$10\%) is not necessary in absolute cross section measurements with the BlueSTEAl 
because the phoswich (with which the bream rate is monitored) and the Si detectors experience the same dead time, and therefore the live time factors cancel. 
This is one of the major benefits in performing absolute cross section measurements with BlueSTEAl. 
As shown in Figure~\ref{fig:Fig12}, the results from the Config. 1 and 2 agree with each other well. 
Cross sections were also calculated with FRESCO \cite{Thompson1988} using optical models for comparison (slightly modified from \cite{Sahm1986,Colomer2016}) and plotted together. 
The theory calculations reproduce the experimental results well. 
The measured cross sections at forward angles ($\sim$10$^{\circ}$ in the CM system ($\theta_{CM}$)) are relatively less well-reproduced by theory. 
It is likely because of the beam's spatial spread. 
Further investigation and data analysis are under way and will be reported in a forthcoming publication. 

% \begin{figure}[!ht]
%        \centering
%          \includegraphics[width=8cm]{figures/SimulatedCrossSections.png}          
%        \caption{(top panel) $^{11}$Be+$^{12}$C angular differential elastic scattering cross sections deduced from the simulated events. (bottom) Deviation of 
%        the deduced cross sections from the theoretical values.  
%\cite{Glatz1986}. %
%        }\label{fig:Fig13}
%    \end{minipage}
%    \hfill{}
%\end{figure}

\section{Future experiments with rare isotope beams}

Following the successful experiment on $^{11}$Be elastic scattering, 
further experiments using more exotic halo nuclei such as $^{19}$C and $^{31}$Ne \cite{Nakamura1999,Nakamura2009} may be expected. 
Moderately intense beam of these nuclei can be produced at FRIB (Facility for Rare Isotope Beams) \cite{FRIBBeam_2023}. 
BlueSTEAl's PID capabilities for these beams are briefly studied using Monte Carlo simulations, in the same manner as described in the previous section, following beam profile at FRIB. 
For each of $^{19}$C and $^{31}$Ne beam, 40 and 45 MeV/u (1.5 and 2.5 MeV/u rms) were assumed (the particles stop in the last Si detector at these energies) with 1 cm (rms) beam spread in $x$ and $y$ directions, respectively \cite{FRIB2023}. 
The four Si detectors were bundled to stop the high energy elastic scattering and elastic neutron breakup events for PID by the $\Delta E-E$, 
and placed at 6$^{\prime\prime}$ downstream from the C target (1 mm thickness). 

\begin{figure}[!ht]
        \centering
          \includegraphics[width=8cm]{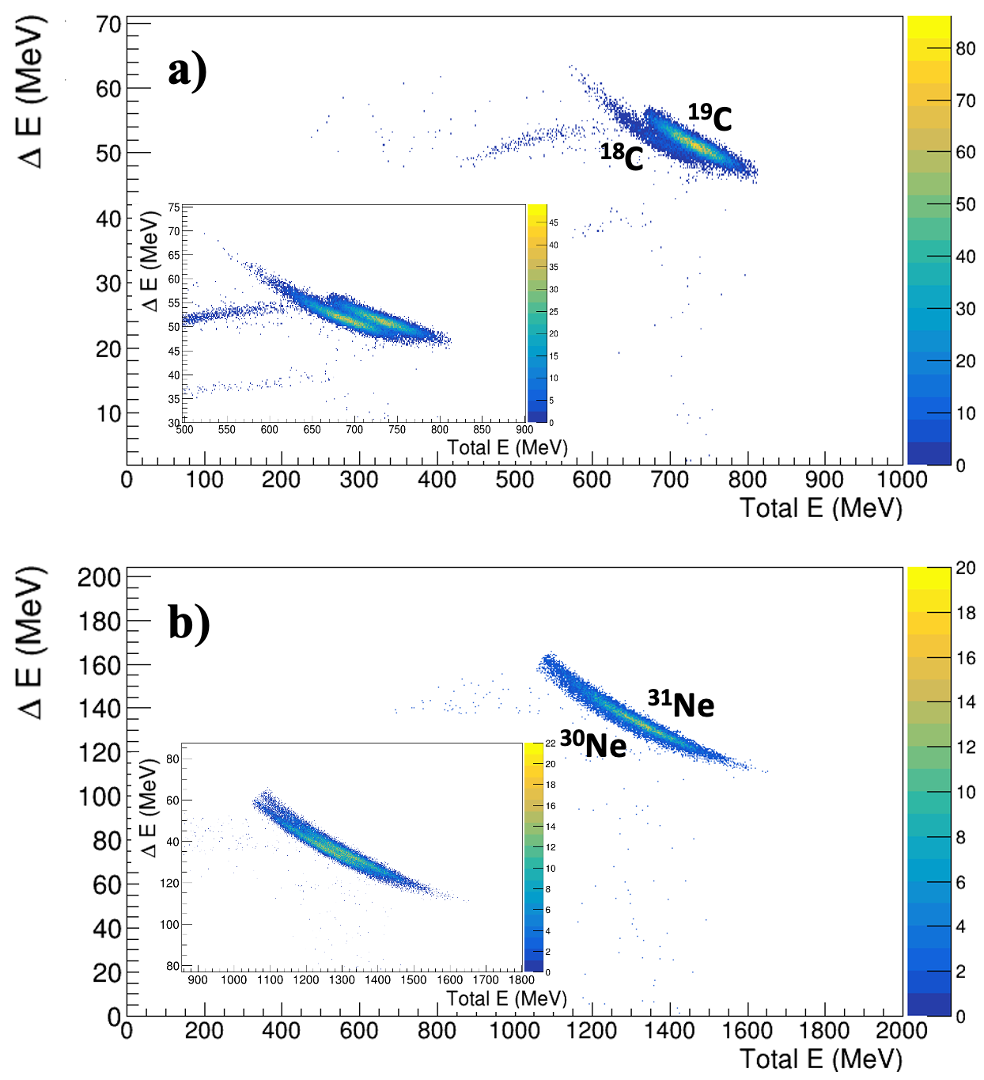}          
        \caption{Simulated particle identification plots by BlueSTEAl Si array, assuming a) 40 MeV/u (rms 1.5 MeV) $^{19}$C and b) 45 MeV/u (rms 2.5 MeV) $^{31}$Ne beam. 10\% statistics were assumed for neutron breakup production. The close-up view is assuming the same statistics for scattering and breakup events for clarity. 
%\cite{Glatz1986}. %
        }\label{fig:Fig14}
%    \end{minipage}
    \hfill{}
\end{figure}

Figure~\ref{fig:Fig14} a) shows PID plots of elastic $^{19}$C and breakup $^{18}$C.  
$^{18}$C is produced assuming elastic one neutron breakup from $^{19}$C+$^{12}$C reaction. 
The statistics for the detected $^{19}$C are $\sim$10$^4$ events which is realistic with several 10$^3$ pps beam after a few days of run. 
The statistics for $^{18}$C is limited to 10\% of $^{19}$C elastic scattering to simulate a realistic situation. 
In the inset plot, for clarity, the case where both particles have the same statistics is considered and the simulated results are shown. 
The isotope separation of these isotopes is clear to perform the elastic and breakup measurements, regardless of statistics. 

Figure~\ref{fig:Fig14} b) shows PID plots of $^{31}$Ne beam and $^{30}$Ne. 
$^{30}$Ne is produced assuming elastic one neutron breakup from $^{31}$Ne+$^{12}$C reaction. 
The plots were made in a similar manner to the $^{19}$C case. 
It turns out that it is difficult to separate these isotopes with BlueSTEAl Si detectors, 
especially when the breakup product is much weaker than elastic/inelastic scattering events. 

From these simulations, we can safely say that 
BlueSTEAl is able to clearly identify rare isotopes of mass ($A$)$<$20. 
To study heavier isotopes, 
improvement to the PID is necessary. 
It should also be noted that the phoswich can easily identify $Z$$\sim$10 as well as demonstrated above, 
and can be used as a beam monitor for these experiments if necessary.

\section{Summary}

BlueSTEAl, the Blue (aluminum chamber of) Silicon TElescope Arrays for light nuclei, has been developed to study direct reactions in inverse kinematics and scattering and breakup reactions using radioactive ion beams. 
It is a detector system consisting of a pair of annular silicon detector arrays and a zero-degree phoswich plastic scintillator, supplemented with some supporting detectors. 
The new detector system, coupled with the new fully-digitized data acquisition (DAQ) system, demonstrated the measurements of events with small cross sections in a high beam flux environment with stable beams. 
We also demonstrated the feasibility of measuring absolute elastic and neutron breakup reaction cross sections using radioactive ion beams ($A<$20) with BlueSTEAl. 
These performance metrics make the device well suited to studying binary reactions in inverse kinematics at Texas A\&M University and other stable beam facilities.
The BlueSTEAl's capability will also enable the future study of exotic nuclei in the era of intense radioactive ion beam experiments at FRIB as well. 
In the latter case, for example, coupling with neutron detectors may make the BlueSTEAl detector system more effective for studying physics near the neutron drip-line using rare isotope beams. Furthermore, implementing a beam tracking device such as the diamond detectors and PPACs with the BlueSteAl is important to improve the accuracy of the cross section measurements, especially at forward angles. Operating the BlueSteAl with these devices is in preparation. 

%There are various bibliography styles available. You can select the style of your choice in the preamble of this document. These styles are Elsevier styles based on standard styles like Harvard and Vancouver. Please use Bib\TeX\ to generate your bibliography and include DOIs whenever available.
%Here are two sample references: \cite{Feynman1963118,Dirac1953888}.

\section*{Ackonwledgement}

%\begin{acknowledgments}
We express our thanks to the technical staff at the Texas A\&M University Cyclotron Institute. 
SO would like to thank Dr. E. A. McCutchan for her helpful comments. 
Financial support for this work was provided by the US Department of Energy, award Nos.\ DE-FG02-93ER40773 and DE-SC0018980, the US National Nuclear Security Administration, award No.\ DE-NA0003841.  
SO acknowledges support from the Office of Nuclear Physics, Office of Science of the U.S. Department of Energy under Contract No.DE-AC02-98CH10886 with Brookhaven Science Associates, LLC. 
GC acknowledges support from the National Sciences and Engineering Research Council of Canada, award  
SAPIN-2020-00052.
WNC, DTD, and GL acknowledge support from the UK STFC, award no.\ ST/L005743/1. 

\bibliography{mybibfile}

\end{document}